\documentclass{aastex}
\usepackage{lscape}
\usepackage{graphicx}
\usepackage{epstopdf}
\usepackage{graphicx}
\usepackage{multirow}
\usepackage{color}

\newcommand{\htwoo}{H$_{2}$O}

\shorttitle{LRS stellar spectra}
\shortauthors{M. G. Kim et al.}

\begin{document}

\title{Low-Resolution Near-infrared Stellar Spectra Observed by the
  Cosmic Infrared Background Experiment (CIBER)}

\author{
Min Gyu Kim\altaffilmark{1,2}, 
Hyung Mok Lee\altaffilmark{1}, 
Toshiaki Arai\altaffilmark{3},
James Bock\altaffilmark{4,5},
Asantha Cooray\altaffilmark{6},
Woong-Seob Jeong\altaffilmark{2},
Seong Jin Kim\altaffilmark{2},
Phillip Korngut\altaffilmark{4,5},
Alicia Lanz\altaffilmark{4},
Dae Hee Lee\altaffilmark{2},
Myung Gyoon Lee\altaffilmark{1},
Toshio Matsumoto\altaffilmark{3},
Shuji  Matsuura\altaffilmark{3,7},
Uk Won Nam\altaffilmark{2},
Yosuke Onishi\altaffilmark{3,8},
Mai Shirahata\altaffilmark{3},
Joseph Smidt\altaffilmark{6,9},
Kohji Tsumura\altaffilmark{10},
Issei Yamamura\altaffilmark{3},
and  
Michael Zemcov\altaffilmark{11,5} 
}

\email{mgkim@astro.snu.ac.kr}

\altaffiltext{1}
{Dept. of Physics and Astronomy, Seoul National University, Seoul 08826, Korea
}
\altaffiltext{2}
{Korea Astronomy and Space Science Institute (KASI), Daejeon 34055, Korea
}
\altaffiltext{3}
{Department of Space Astronomy and Astrophysics, Institute of Space and Astronautical Science (ISAS), Japan Aerospace Exploration Agency (JAXA), 3-1-1 Yoshinodai, Chuo-ku, Sagamihara, Kanagawa 252-5210, Japan
}
\altaffiltext{4}
{Department of Astronomy, California Institute of Technology, Pasadena, CA 91125, USA
}
\altaffiltext{5}
{Jet Propulsion Laboratory (JPL), 4800 Oak Grove Dr., Pasadena, CA 91109, USA
}
\altaffiltext{6}
{Center for Cosmology, University of California, Irvine, Irvine, CA 92697, USA
}
\altaffiltext{7}
{Department of Physics, Kwansei Gakuin University, Hyogo 669-1337, Japan
}
\altaffiltext{8}
{Department of Physics, Tokyo Institute of Technology 2-12-1 Ookayama, Meguro-ku, Tokyo, 152-8550, Japan
}
\altaffiltext{9}
{Theoretical Division, Los Alamos National Laboratory, Los Alamos, NM 87545, USA
}
\altaffiltext{10}
{Frontier Research Institute for Interdisciplinary Science, Tohoku University, Sendai 980-8578, Japan
}
\altaffiltext{11}
{Center for Detectors, School of Physics and Astronomy, Rochester Institute of Technology, Rochester NY 14623, USA
}

\begin{abstract}
  We present near-infrared (0.8-1.8 $\micron$) spectra of 105 bright
  (${m_{J}}$ $<$ 10) stars observed with the low resolution spectrometer
  on the rocket-borne Cosmic Infrared Background Experiment
  (CIBER). As our observations are performed above
  the earth's atmosphere, our spectra are free from telluric
  contamination, which makes them a unique resource for near-infrared spectral calibration. 
  Two-Micron All Sky Survey (2MASS) photometry information is
  used to identify cross-matched stars after reduction and extraction
  of the spectra. We identify the spectral types of the observed stars by
  comparing them with spectral templates from the Infrared Telescope
  Facility (IRTF) library. All the observed spectra are consistent
  with late F to M stellar spectral types, and we identify various
  infrared absorption lines. 
\end{abstract}

\keywords{catalogs --- infrared: stars --- stars: general --- techniques: spectroscopic}

\section{ Introduction }

Precise ground-based measurements of stellar spectra are challenging
in the near-infrared (IR) because of the contaminating effects of
telluric lines from species like water, oxygen, and hydroxyl in the
earth's atmosphere. Telluric correction using standard stars is
generally used to overcome this problem, but these corrections are
problematic in wavelength regions marked by strong line
contamination, such as from water and hydroxyl.
In contrast, space-based spectroscopy in the
near-IR does not require telluric correction, so can provide new
insights into stellar atmospheres (e.g.~\citealt{matsuura99,tsuji01}),
especially near $1 \micron$ where starlight is not reprocessed by dust
in the circumstellar environment \citep{meyer98}. In particular,
near-IR spectra can be used to study the age and mass of very young
stars \citep{joyce98,peterson08}, and the physical properties of very
cool stars \citep{sora14}.

Of particular interest in the study of the atmospheres of cool stars
is water.  According to early models of stellar photospheres \citep{russell34}, \htwoo\
existed only in later than M6 type stars, and until recently observations have
supported this.
In 1963, the balloon-borne telescope Stratoscope II observed \htwoo\ in two early M2-M4 giant
stars \citep{woolf64} at 1.4 and $1.9 \, \micron$. Several decades
later, \citet{tsuji97} measured \htwoo\ absorption in an M2.5 giant
star using the Infrared Space Observatory (\citealt{kessler96}),
and \citet{matsuura99} observed water at 1.4, 1.9, 2.7, and
$6.2 \micron$ for 67 stars with the Infrared Telescope in Space (\citealt{murakami96,matsumoto05}).
Surprisingly, \citet{tsuji01} discovered
water features in late K-type stars.  These results required a new
stellar photosphere model to explain the existence of \htwoo\ features
in hotter than M6 type stars \citep{tsuji15}.

The low resolution spectrometer (LRS; \citealt{tsumura13}) on the
Cosmic Infrared Background Experiment (CIBER;
\citealt{bock06,zemcov13}) observed the diffuse infrared background
from 0.7 to 2.0 $\micron$ during four flights above the Earth
atmosphere. The LRS was designed to observe the near-IR background
\citep{hauser01,madau00}, and as a result finds excess
extragalactic background light above all known foregrounds \citep{matsuura16}.
Furthermore, we precisely measure astrophysical components
contributing to the diffuse sky brightness (see \citealt{leinert98} for a review).
For example, \citet{tsumura10} observed a component of the zodiacal light
absorbed by silicates in a broad band near $800 \,$nm.  By correlating the LRS with a 100 $\micron$
dust map \citep{schlegel98}, \citet{arai15} measured smooth diffuse
galactic light spectrum from the optical band to the near-IR and constrained
the size distribution of interstellar dust, which was dominated by small
particles (half-mass radius $\sim$0.06 $\micron$).  

The LRS also observed many bright galactic stars, enabling us to study
their near-IR SEDs.  In this paper, we present flux-calibrated near-IR
spectra of 105 stars from $0.8 \leq \lambda \leq 1.8 \, \micron$ with
spectral resolution $15 \leq \lambda / \Delta \lambda \leq 30$ over
the range.
The paper is organized as follows. In Section \ref{S:instrument},
the observations and instrumentation are introduced. We describe the data
reduction, calibration, astrometry, and extraction of the stellar spectra
in Section \ref{S:datareduction}.  In Section \ref{S:stars}, the spectral
typing and features are discussed.  Finally, a summary and discussion
are given in Section \ref{S:summary}.

\section{Instrument}
\label{S:instrument}

The LRS is one of the four optical instruments of the CIBER payload \citep{zemcov13}; the others are
a narrowband spectrometer (\citealt{korngut13}) and two
wide-field imagers \citep{bock13}.  The LRS \citep{tsumura13} is a
prism-dispersed spectrometer with five rectangular
$5.35^{\circ} \times 2.8 \arcmin$ slits imaging a 5.8 $^{\circ}$
$\times$ 5.8 $^{\circ}$ field of view.  The detector has
$256 \times 256$ pixels at a pixel scale of
$1.36 \arcmin \times 1.36 \arcmin$.  CIBER has flown four times (2009
February, 2010 July, 2012 March, and 2013 June) with apogees and total exposure times of
over 325 km and $\sim$ 240 s, respectively, in the first three
flights and of 550 km and 335 s in the final, non-recovered flight.
Due to spurious signal contamination from thermal emission from the
shock-heated rocket skin, we do not use the first flight data in this
work \citep{zemcov13}.  Eleven target fields were observed during
the three subsequent flights, as listed in Table \ref{tbl1}.
Details of the field selection are described in \citet{matsuura16}.

During the observations, the detector array is read nondestructively
at $\sim 4 \,$Hz frame$^{-1}$.  Each field is observed for many tens or hundreds
of frames, and an image for each field is obtained by computing the
slope of the accumulated values for each pixel \citep[]{gf93}. Figure \ref{flight_image} shows an
example image of the North Ecliptic Pole (NEP) region obtained during the
second flight.  More than 20 bright stars ($m_{J}$ $<$ 11) are
observed.  The stellar spectra are characterized by a small amount of field
distortion as well as an arc-shaped variation in constant-wavelength
lines along the slit direction.  The latter is known as a ``smile'' and
is a known feature of prism spectrometers
\citep{fischer98}. Details of the treatment of these
distortions are described in Section \ref{sS:SBR} and \ref{sS:starsel}.

\section{Data Analysis}
\label{S:datareduction}

In this section, we describe how we perform background subtraction,
calibration, photometric estimation, astrometric registration, and
spectral extraction from the LRS-observed images.  

\subsection{Pixel response correction}
\label{sS:PRC}

We measure the relative pixel response (flat field) in the laboratory
before each flight \citep{arai15}. The second- and the third-flight data are normally corrected with these laboratory flats.
However, for the fourth flight from the laboratory calibrations do not extend to the longest wavelengths ($\lambda \geq 1.4 \micron$)
because the slit mask shifted its position with respect to the detector during the flight.
We therefore use the second-flight flat field to correct the relative
response for the fourth-flight data, as this measurement covers
$\lambda > 1.6 \micron$.  To apply this flat field, we need to assume
that the intrinsic relative pixel response does not vary significantly over the flights.
To check the validity of this assumption, we subtract the second flat image to the fourth flat image
for overlapped pixels and calculate the pixel response difference.
We find that only 0.3 \% of pixels with response measured in both are different by $2 \sigma$,
where $\sigma$ is the standard deviation of the pixel response.
Finally, we mask 0.06 \% of the array detectors to remove those pixels
with known responsivity pathologies and those prone to transient electronic
events \citep{lee10}.

\subsection{Calibration}

For each flight, the absolute brightness and wavelength irradiance
calibrations have been measured in the laboratory in collaboration with
the National Institute of Standards and Technology.
The details of these calibrations can be found in \citet{tsumura13}.
The total photometric uncertainty of the LRS brightness calibration is
estimated to be $\pm 3$\% \citep{tsumura13,arai15}.

\subsection{Background Removal}
\label{sS:SBR}

The raw image contains not only spectrally dispersed images of stars
but also the combined emission from zodiacal light
$\lambda I_{\lambda}^{\rm ZL}$, diffuse galactic light
$\lambda I_{\lambda}^{\rm DGL}$, the extragalactic background
$\lambda I_{\lambda}^{\rm EBL}$, and instrumental effects
$\lambda I_{\lambda}^{\rm inst}$ \citep{leinert98}.  The measured
signal $\lambda I_{\lambda}^{\rm meas}$ can be expressed as
\begin{equation}
\lambda I_{\lambda}^{\rm meas} = \lambda I_{\lambda}^{\ast} + 
\lambda I_{\lambda}^{\rm ZL} + \lambda
I_{\lambda}^{\rm ISL} + \lambda I_{\lambda}^{\rm DGL} + \lambda
I_{\lambda}^{\rm EBL} + \lambda I_{\lambda}^{\rm inst},
\end{equation}
where we have decomposed the intensity from stars into a resolved
component $\lambda I_{\lambda}^{\ast}$ and an unresolved component
arising from the integrated light of stars below the sensitivity of the
LRS $\lambda I_{\lambda}^{\rm ISL}$.  It is important to subtract the
sum of all components except $\lambda I_{\lambda}^{\ast}$ from the
measured brightness to isolate the emission from detected stars.  At
this point in the processing, we have corrected for multiplicative
terms affecting $\lambda I_{\lambda}^{\rm meas}$.  Dark current, which is the detector photocurrent
measured in the absence of incident flux, is an additional contribution to $\lambda I_{\lambda}^{\rm inst}$.
The stability of the dark current in the LRS has been shown to be 0.7 nW m$^{-2}$ sr$^{-1}$ over each flight,
which is a negligible variation from the typical dark current
(i.e., 20 nW m$^{-2}$ sr$^{-1}$; \citep{arai15}).  As a result, we
subtract the dark current as part of the background estimate formed below.

\begin{figure*}[htp]
	\centering \epsscale{0.8} \plotone{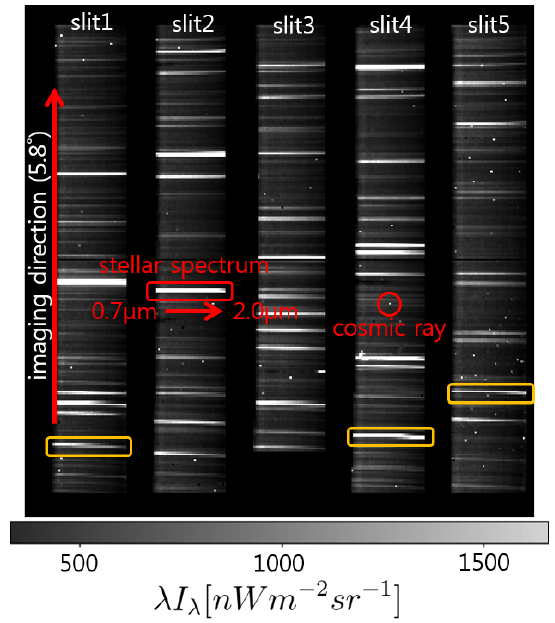}
	\caption{An example CIBER-LRS image toward the NEP field. The five illuminated
		columns are dispersed spectra from the five slits of the
		LRS, and the bright horizontal lines in each column are
		images of individual stars. As an example, we highlight a
		single horizontal light trail by a red box; this is the
		light from a single star dispersed from 0.7 to 2.0 $\micron$.
		The bright dots are pixels hit by cosmic rays.
		The yellow boxes highlight representative
		examples of stellar spectra disturbed by the prism.
		Note that the distortion direction is different between the upper and lower parts of the image,
		and the distortion becomes negligible at the center line of the image. \label{flight_image}
		}
\end{figure*}

The relative brightnesses of the remaining background components are
wavelength-dependent, so an estimate for their mean must be computed
along constant-wavelength regions, corresponding to the vertical columns
in Figure \ref{flight_image}.  Furthermore, because of the LRS's large
spatial PSF, star images can extend over several pixels in the imaging
direction and even overlap one another.  This complicates background
estimation in pixels containing star images and reduces the number of
pixels available to estimate the emission from the background components.

\begin{figure*}[p]
	\centering
	\epsscale{0.85}
	\plotone{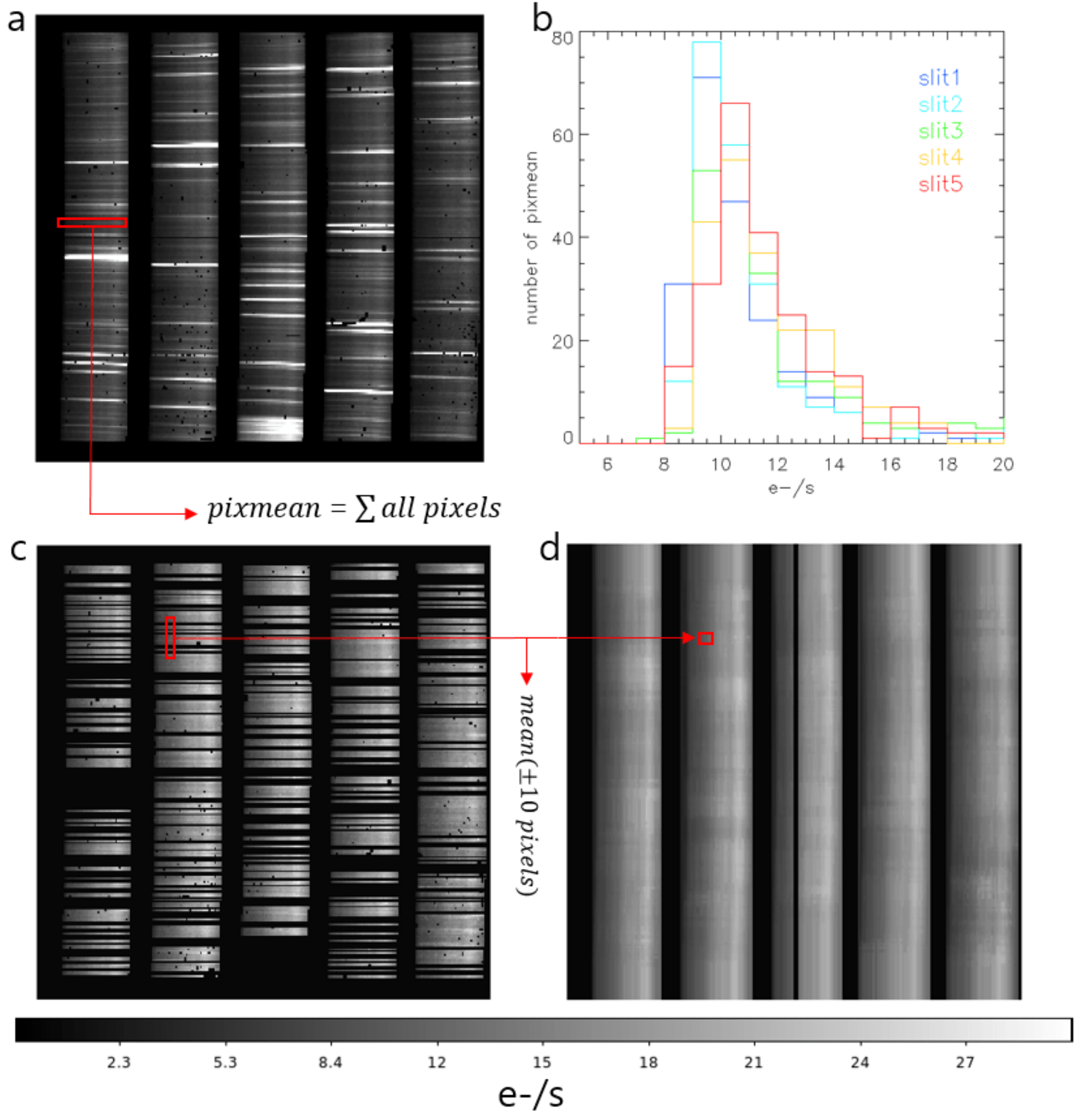}
	\caption{Flow chart of the background image construction. (\textit{a}) Same as Figure \ref{flight_image}. The red box indicates the set of rows to be averaged. (\textit{b}) Histogram of averaged values for each row. This average values for each slit are drawn with different color. (\textit{c}) Image after iterative sigma clipping of bright rows from (\textit{b}). The red box indicates the size of $\pm$ 10 pixels that are averaged. (\textit{d}) Reconstructed background image including all instrumental noise and undetected faint stars.\label{flow_chart}}
\end{figure*}

To estimate the background in those pixels containing star images,
we compute the average value of pixels with no star images along each
column, as summarized in Figure \ref{flow_chart}. 
We remove bright pixels that may contain star images, as described in \citet{arai15}.
The spectral smile
effect shown in Figure \ref{flight_image} introduces spectral
curvature along a column. We estimate it causes an error of magnitude
$\delta \lambda / \lambda < 10^{-2}$, which is small compared to the
spectral width of a pixel. Approximately half of the rows remain
after this clipping process; the fraction ranges from 45 \% to 62 \%
depending on the stellar density in each field.  This procedure
removes all stars with $J> 13$, and has a decreasing completeness
above this magnitude \citep{arai15}.

To generate an interpolated background map,
each candidate star pixel is replaced by the average of nearby pixels
calculated along the imaging direction from the $\pm 10$ pixels on
either side of the star image.  We again do not explicitly account for
the spectral smile.
This interpolated background image is subtracted from the measured
image, resulting in an image containing only bright stellar emission.
The emission from faint stars and bright stars that inefficiently
illuminate a grating slit that contributes to $I_{\lambda}^{\rm ISL}$
is naturally removed in this process.

\subsection{Star Selection}
\label{sS:starsel}

The bright lines dispersed in the spectral direction in the
background-subtracted images are candidate star spectra.  To calculate
the spectrum of candidate sources, we simply isolate individual
lines of emission and map the pixel values onto the wavelength using the
ground calibration.  However, this procedure is complicated both by
the extended spatial PSF of the LRS and by source confusion.

To account for the size of the LRS spatial PSF (FWHM $\sim$1.2 pixels) as
well as optical distortion from the prism that spreads the star images
slightly into the imaging direction, we sum five rows of pixels in the
imaging direction for each candidate star. Since the background
emission has already been accounted for, this sum converges to the total
flux as the number of summed rows is increased.  By summing five rows, we
capture $>99.9$\% of a candidate star's flux. The wavelengths of the
spectral bins are calculated from the corresponding wavelength
calibration map in the same way.

From these spectra, we can compute synthetic magnitudes in the $J$- and
$H$-bands, which facilitate comparison to Two-Micron All-Sky Survey (2MASS) measurements.  We
first convert surface brightness in nW m$^{-2}$ sr$^{-1}$ to flux in
nW m$^{-2}$ Hz$^{-1}$, and then integrate the monochromatic intensity
over the 2MASS band, applying the filter transmissivity of the $J$- and
$H$-bands \citep{cohen03}.  To determine the appropriate zero
magnitude, we integrate the $J$- and $H$-band intensity of Vega's
spectrum \citep[]{bohlin04} with the same filter response. The $J$- and
$H$-band magnitudes of each source are then calculated, allowing both
flux and color comparisons between our data and the 2MASS catalog.

Candidate star spectra may be comprised of the blended emission from two or
more stars, and these must be rejected from the catalog. Such blends
fall into one of two categories: (\textit{i}) stars that are visually
separate but are close enough to share flux in a 5 pixel-wide
photometric aperture, and (\textit{ii}) stars that are close enough
that their images overlap so as to be indistinguishable.  We isolate
instances of case (\textit{i}) by comparing the fluxes calculated by
summing both three and five rows along the imaging direction for each source.
If the magnitude or $J - H$ color difference between the two apertures
is larger than the statistical uncertainty (described in Section
\ref{sS:errors}), we remove those spectra from the catalog.  To find
instances of case (\textit{ii}), we use the 2MASS star catalog
registered to our images using the procedure described in Section
\ref{sS:astrometry}. Candidate sources that do not meet the criteria
presented below are rejected.

\begin{figure*}[p]
	\centering
	\epsscale{1}
	\plotone{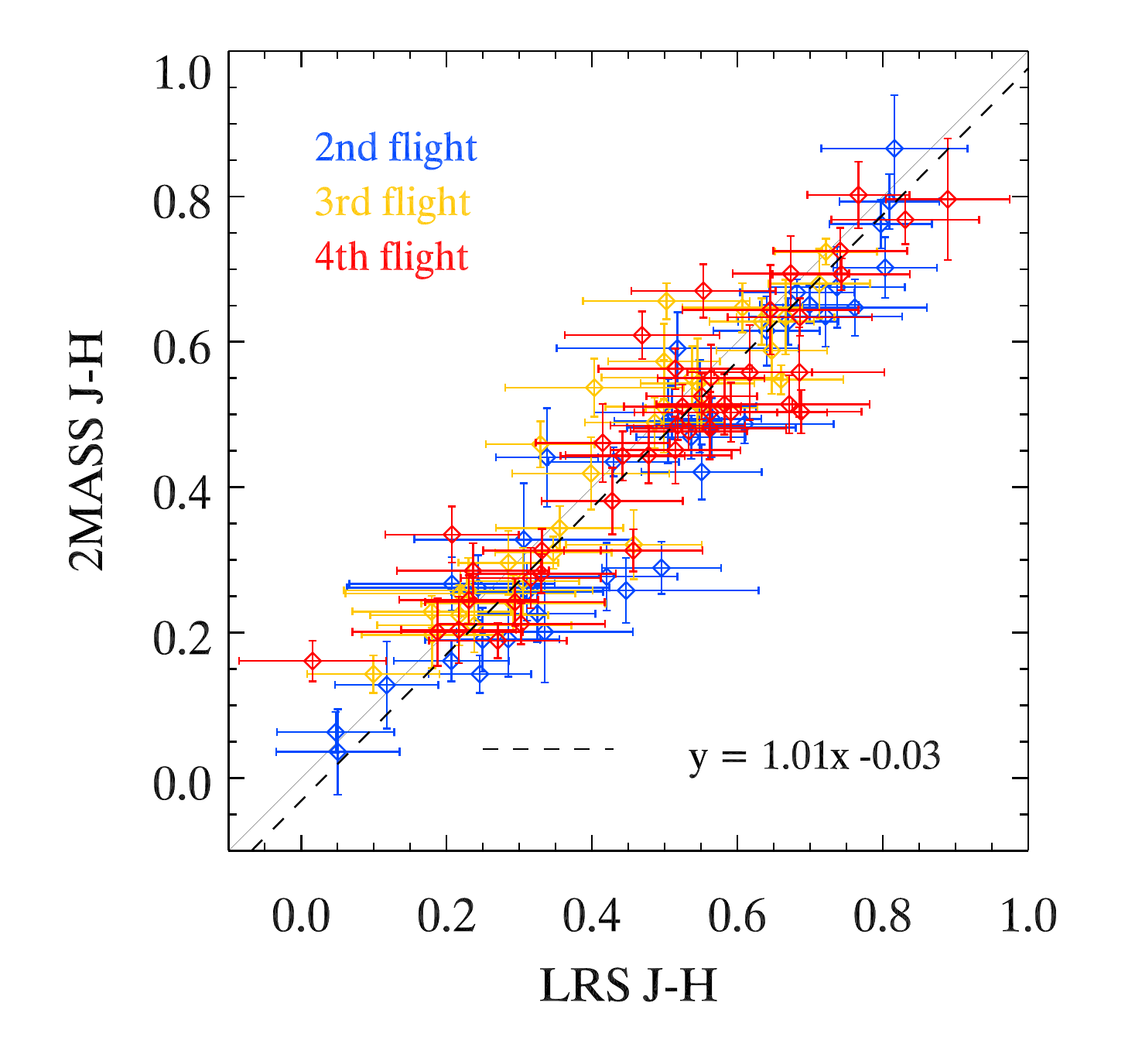}
	\caption{LRS J-H color comparison with cross-matched 2MASS J-H
          color. Each color corresponds to a different flight. The dashed line shows a
          linear fit, exhibiting a slight systematic offset from unity.
          The J-H colors of LRS stars are conserved regardless of the slit apodization effect.\label{color_comp}}
\end{figure*}

To ensure the catalog spectra are for isolated stars rather than
for indistinguishable blends, we impose the following requirements on
candidate star spectra: (\textit{i}) each candidate must have $J< 11$;
(\textit{ii}) the $J$-band magnitude difference between the LRS
candidate and the matched 2MASS counterpart must be $<1.5$;
(\textit{iii}) the $J - H$ color difference between the LRS candidate star and the
matched 2MASS counterpart must be $< 0.3$; and (\textit{iv}) among the
candidate 2MASS counterparts within the 500$\arcsec$ ($=6$ pixel) radius
of a given LRS star, the second-brightest 2MASS star must be fainter
than the brightest one by more than 2 mag at the J band. Criterion
(\textit{i}) excludes faint stars that may be strongly affected by
residual backgrounds, slit mask apodization, or source confusion. The
second and third criteria mitigate mismatching by placing requirements
on the magnitude and color of each star.  In particular, the $J-H$
color of a source does not depend on the slit apodization or the position
in image space (see Figure \ref{color_comp}), so any significant
change in $J-H$ color as the photometric aperture is varied suggests
that more than a single star could be contributing to the measured
brightness.  Finally, it is possible that two stars with similar $J-H$
colors lie close to each other, so the last criterion is applied to
remove stars for which equal-brightness blending is an issue.
Approximately one in three candidate stars fails criterion
(\textit{iv}). The number of candidate stars rejected at each criterion is described in Table \ref{tbl3}.

In addition, three of LRS candidate stars are identified as variables in the
SIMBAD database \footnote{http://simbad.u-strasbg.fr/simbad/}.
We also identify two stars as binary and multiple-star systems as well as four high proper motion stars.
Through these stringent selection requirements, we conservatively include only the spectra of bright,
isolated stars in our catalog. Finally, 105 star spectra survive all the cuts, and the corresponding stars
are selected as catalog members.

\subsection{Astrometry}
\label{sS:astrometry}

We match the synthesized LRS $J$, $H$, and
$J-H$ information with the 2MASS point source catalog
\citep{skrutskie06} to compute an astrometric solution for the LRS
pointing in each sky image.  This is performed in a stepwise fashion
by using initial estimates for the LRS's pointing to solve for
image registration on a fine scale.

As a rough guess at the LRS pointing, we use information provided by
the rocket's attitude control system (ACS), which controls the pointing
of the telescopes \citep{zemcov13}. This provides an estimated pointing
solution that is accurate within 15 $\arcmin$ of the requested coordinates.
However, since the ACS and the LRS are not explicitly aligned to one
another, a finer astrometric registration is required to capture the
pointing of the LRS to single-pixel accuracy.

To build a finer astrometric solution, we simulate images of each
field in the 2MASS \textit{J}-band using the positional information from the
ACS, spatially convolved to the LRS PSF size. Next, we apodize
these simulated 2MASS images with the LRS slit mask, compute the
slit-masked magnitudes of three reference stars, and calculate the
$\chi^2$ statistic using
\begin{equation}
	\chi^2{_{p,q}}=\sum_{i}{}^{} \left (
	\frac{F_{LRS,i}-F_{2MASS,i}}{\sigma_{LRS,i}} \right
	)^2, 
\end{equation}
where index \textit{i} represents each reference star and subscripts
\textit{p} and \textit{q} index the horizontal and vertical positions of the slit mask,
respectively. $F_{LRS,i}$ and $F_{2MASS,i}$ are the fluxes in the LRS and
2MASS $J$-band, and $\sigma_{LRS,i}$ is the statistical error of
the LRS star (see Section \ref{sS:errors}). The minimum $\chi^2$ gives the
most likely astrometric position of the slit mask. Since, on
average, there are around five bright stars with $J<9$ per field,
spurious solutions are exceedingly unlikely, and all fields give a
unique solution. 

\begin{figure*}[p]
	\centering
	\epsscale{1}
	\plotone{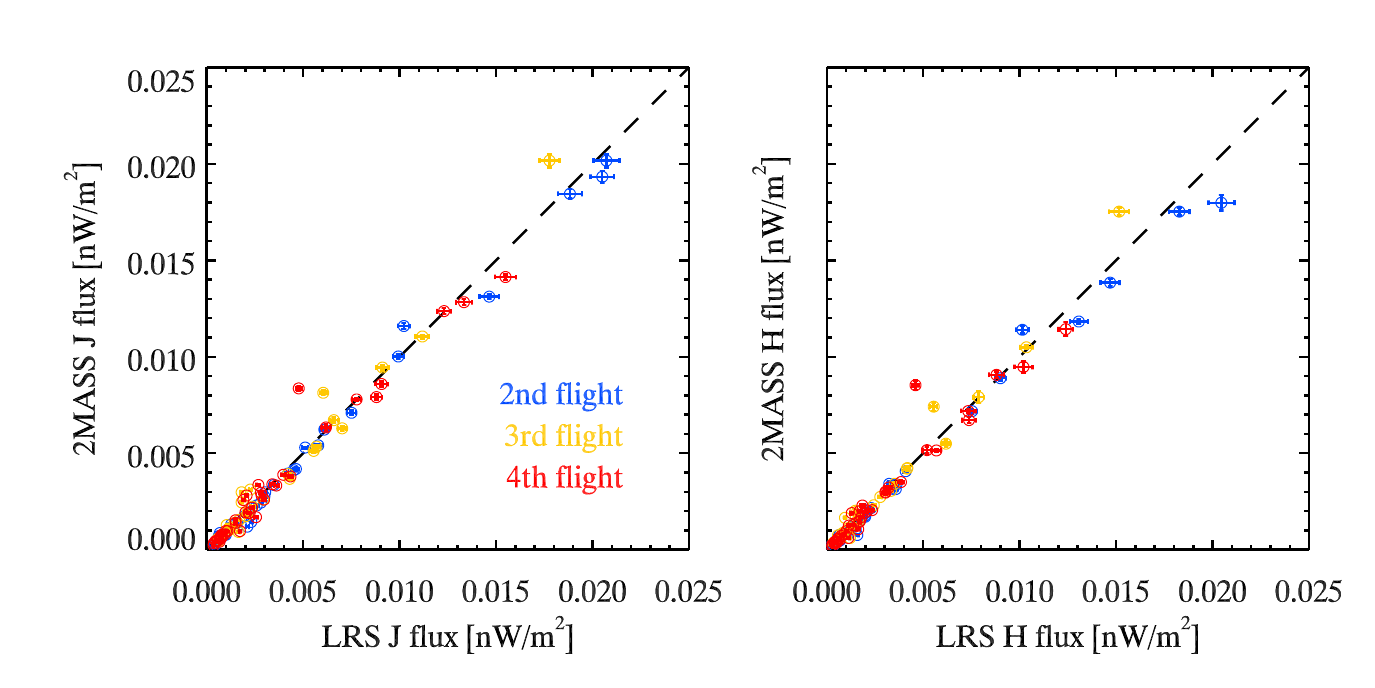}
	\caption{The 2MASS J- and H-band fluxes are shown as a function of the LRS J- and H-band. Each color represents the data obtained on a different flight. Slit apodization effect is corrected for all LRS stars. Correction factors are derived based on the slit simulation for magnitude ranges covered by the LRS stars, as shown in Figure \ref{model_J} and \ref{model_H}.\label{flux_comp}}
\end{figure*}

Using this astrometric solution, we can assign coordinates to the rest
of the detected LRS stars. We estimate that the overall astrometric
error is 120$\arcsec$ by computing the mean distance between the LRS and
2MASS coordinates of all matched stars.  The error corresponds to 1.5
times the pixel scale. We check the validity of the astrometric solutions
by comparing the colors and fluxes between the LRS and matched 2MASS stars.
In Figures \ref{color_comp} and \ref{flux_comp}, we show the comparison
of the $J-H$ colors and fluxes of the cross-matched stars in each
field. Here, we multiply the LRS fluxes at the J- and H-band by 2.22 and 2.17, respectively, to 
correct for the slit apodization. The derivation of correction factors is described in Section \ref{S:summary}.
On the whole, they match well within the error range.

\subsection{Spectral Error Estimation}
\label{sS:errors}

Even following careful selection, the star spectra are subject to
various kinds of uncertainties and errors, including statistical
uncertainties, errors in the relative pixel response, absolute
calibration errors, wavelength calibration errors, and background
subtraction errors.

Statistical uncertainties in the spectra can be estimated directly
from the flight data.  We calculate the $1 \sigma$ slope error from
the line fit (see Section \ref{S:instrument}) as we generate the flight
images; this error constitutes the estimate for the statistical photometric
uncertainty for each pixel. In this statistical error, we include
contributions from the statistical error in the background estimate
and the relative pixel response. The error in the background signal
estimate is formed by computing the standard deviation of the $\pm$10
pixels along the constant-$\lambda$ direction for each pixel to match
the background estimate region. This procedure captures the local
structure in the background image, which is a reasonable measure of
the variation we might expect over a photometric aperture. Neighboring
pixels in the wavelength direction have extremely covariant error
estimates in this formulation, which are acceptable since the flux
measurements are also covariant in this direction.
A statistical error from the relative
pixel response correction is applied by multiplying 3\% of the relative
response by the measured flux in each field \citep{arai15}.
To compute the
total statistical error, each constituent error is summed in
quadrature for each pixel.

Several instrumental systematic errors are present in these
measurements, including those from wavelength calibration, absolute calibration,
and relative response correction. In this work, we do not
explicitly account for errors in the wavelength calibration, as the
variation is $\pm$ 1 nm over 10 constant-wavelength pixels, which is
$< 0.1 R$.  In all flights, $<$ 3 \% absolute calibration error is
applied \citep{arai15}.
For the longest-wavelength regions ($\lambda$ $>$ 1.6 $\micron$) of the fourth-flight data that are not measured even in the second-flight flat, we could not perform flat correction. Instead, we apply a systematic error amounting to 5.3 \% of the measured sky brightness. The error is estimated from pixels in the short-wavelength regions ($\lambda$ $<$ 1.4 $\micron$) of the fourth-flight flat. We calculate deviations from unity for those pixels and take a mean of 5.3 \%.
The linear sum of systematic errors is then combined with statistical error in quadrature.

\section{The Spectra}
\label{S:stars}

The 105 stellar spectra that result from this processing can be used to test
spectral type determination algorithms and study near-IR features that
are invisible from the ground.
Despite the relatively low spectral resolution of our stellar spectra,
we identify several molecular bands, particularly for the late-type
stars. We present the $J-$band-normalized LRS spectra for each of the catalog stars in
Figure \ref{sp_fig1}. 

General information for each
spectrum is summarized in Table \ref{tbl2} with the corresponding star ID.
All spectra are publicly available in electronic form
\footnote{http://astro.snu.ac.kr/$\sim$mgkim/}.
The spectra are presented without the application of interstellar
extinction corrections, since extinction correction assumes both a
color index and the integrated Galactic extinction along the
line of sight. Therefore, without knowing the stars' distances, it is difficult to make progress.
For CIBER fields, typical extinction ranges from 0.005 to 0.036 mag at the J-band
if we assume extinction coefficients R(J) with 0.72 \citep{yuan13}

\subsection{Spectral type determination}

The star spectral types are determined by fitting known spectral
templates to the measured LRS spectra.  We use the Infrared Telescope Facility (IRTF) and Pickles
\citep[]{pickles98} templates for the SED fitting. 
The SpeX instrument installed on the IRTF
observed stars using a medium-resolution spectrograph (\textit{R} $=$ 2000).
The template library contains spectra for 210 cool stars (F to M type) with wavelength coverage
from 0.8 to 2.5 $\micron$ \citep[]{cushing05,rayner09}.
The Pickles library is a synthetic spectral library
that combines spectral data from various observations to achieve
wavelength coverage from the UV (0.115 $\micron$) to the near-IR (2.5
$\micron$). It contains 131 spectral templates for all star types
(i.e., O to M type) with a uniform sampling interval of 5 $\AA$.

To perform the SED fit, we degrade the template spectra to the LRS
spectral resolution using a mean box-car smoothing kernel
corresponding to the slit function of the LRS. Both the measured and template
spectra are normalized to the $J$-band flux. We calculate the flux
differences between the LRS and template spectra using
\begin{equation}
\chi^2=\sum_{\lambda}{}^{} \left (
  \frac{F_{LRS,\lambda}-F_{ref,\lambda}}{\sigma_{LRS,\lambda}} \right
)^2, 
\end{equation}
where $F_{LRS,\lambda}$ and $F_{ref,\lambda}$ are the fluxes of the
observed and template spectra at wavelength $\lambda$ normalized at
$J$-band and $\sigma_{LRS,\lambda}$ is the statistical error of the
observed spectrum. The best-fitting spectral type is determined by
finding the minimum $\chi^2$.

No early-type (i.e., O, B, A) stars are found in our sample; all stars have characteristics
consistent with those of late-type stars (F and later). Because the IRTF library
has about twice the spectral type resolution of the Pickles library,
we provide the spectral type determined from the IRTF
template in Table \ref{tbl2}.  Since the IRTF library does not include a continuous set of spectral
templates, we observe discrepancies between the LRS and best-fit
IRTF templates, even though the $J-H$ colors are consistent between
2MASS and the LRS within the uncertainties.
The Pickles and IRTF fits are consistent within the uncertainty in the
classification ($\sim$ 0.42 spectral subtypes). 

A color-color diagram for the star sample is shown in Figure \ref{ccd}.
Although the color-color diagram does not allow us to clearly discriminate between
spectral types, qualitatively earlier-type stars are located in
the bluer region, while later-type stars are located in the redder
region, consistent with expectations. LRS stars well follow the color-color distributions of
typical 2MASS stars in LRS fields, as indicated by the gray dots.

To estimate the error in our spectral type determination, we compare
our identifications with the SIMBAD database \citep[]{wenger00}, where
63 of the 105 stars have prior spectral type determinations. Figure
\ref{type_difference} shows the spectral types determined from the IRTF fit
versus those from the SIMBAD database.
The 1$\sigma$ error of type difference is estimated to be
0.59 spectral subtypes, which is comparable with those in other published works
\citep[]{gliese71,jaschek73,jas78,roeser88,houk99}. The error can be explained with
two factors: (\textit{i}) the low spectral resolution of the LRS and
(\textit{ii}) the SED template libraries, which do not represent all
star types.

    Five stars are observed twice in different flights (BA2\_5 and
	BB4\_6, N2\_6 and N3\_5, BA2\_1 and BA3\_4, BB2\_1 and BB3\_1, and BB2\_4 and BB3\_4; see Figure \ref{dup_star}),
	enabling us to investigate the interflight stability of the
	spectra.  For BA2\_5 and BB4\_6, the spectral type is known to be F8,
	while our procedure yields F7V and F1II from the
	second- and fourth-flight data, respectively. For N2\_6 and N3\_5, the known type is
	K5 while we determine M0.5V for both flights.
	For BA2\_1 and BA3\_4, the known
	type is F5 while we determine F7III and F2III-IV in the second and third
	flights. For BB2\_1 and BB3\_1, the fitted types are G8IIIFe5 and K4V for a K1 type star,
	and the type of BB2\_4 and BB3\_4 are not known but are fitted to F9V for both flights.
The determined spectra are consistent within an acceptable
error window, though the longer-wavelength data exhibit large
differences, which can be attributed to calibration error.
We present the spectra of each star from both flights in Table
\ref{tbl2}. This duplication results in our reporting of 110 spectra in
the catalog, even through only 105 individual stars are observed.

\section{Discussion}
\label{S:summary}

We determined the spectral type of 105 stars as well as the associated
typing error (0.59 spectral subtypes) assessed by comparing the type against a
set of 63 previously determined spectral types.
Representative examples of the measured spectra for different spectral
types are shown in Figure \ref{type_example}. Molecular absorption
lines are evident in these spectra, including the CaII triplet and
various CN bands.

Since we observed stars above the earth's atmosphere, observations of
the H$_{2}$O molecular band are possible.  However, they are not able
to distinguish between CN and H$_{2}$O at 1.4 $\micron$ since both
have the same bandhead and appear in late-type stars \citep{ws70}.
For example, the spectral features of M2-M4 (super)giant stars observed by Stratoscope II,
previously identified as CN, were identified as H$_{2}$O \citep{tsuji00}.
Several subsequent observations show clear evidence that water features exist
even in K type stars, requiring modifications of present stellar photosphere models \citep{tsuji00}.

In our spectral catalog, most K and M type stars
exhibit a broad absorption band around 1.4 $\micron$.
Although it is not possible to identify specific molecular bands with our data,
we cannot exclude the presence of H$_{2}$O in the spectra of these stars.
Future mid-IR measurements at $6.3 \micron$ would help disentangle
the source of the spectral features by removing the spectral degeneracies
between CN and \htwoo\ \citep{tsuji01}.

As these spectra are free from telluric contamination and the LRS is
calibrated against absolute irradiance standards \citep{arai15},
\textit{in principle} these measurements could be used as near-IR spectral standards.
However, our lack of knowledge of the instrument response function (IRF) on the spectral
plane complicates the use of these measurements for the absolute
photometric calibration of stars. Specifically, the LRS's IRF depends on the
end-to-end optical properties of the instrument. Because we use a
slit mask at the focus of an optical coupler \citep{tsumura13}, the
full IRF knowledge of the focusing element of the optical coupler is difficult to
disentangle from other effects. As a result, we would need to know the
precise IRF to assign an absolute error estimate to an absolute
calibration of the star images. This response function was not
characterized during ground testing.

Nevertheless, we consider it instructive to check the validity of photometric results
whether or not the estimated magnitudes of the LRS stars are reasonable compared to previous measurements.
We perform an empirical simulation as follows. For each LRS
star, we generate a point source image with the flux of the 2MASS
counterpart convolved to the LRS PSF. Instrumental noise and source
confusion from faint stars ($J> 13$) based on the 2MASS stars around a
target star are also added.  We measure the photometric flux of the
simulated star image in the same way as for the LRS stars as described in
this paper. An aperture correction is applied to the LRS stars,
since stars that are clipped by the slit mask will appear to have a reduced flux
measurement. Figures \ref{model_J} and \ref{model_H} show the ratios
of the band-synthesized flux of each LRS star to the flux of the
corresponding 2MASS star with statistical errors.
The range explained by our simulations is illustrated as a color-shaded area.
The LRS stars fall within the expected flux range.
Also, the flux ratios of the stars between flights well agree,
validating the stability of the photometric calibrations for the three CIBER flights.
The large scatter at faint stars is caused by background noise, including adjacent faint stars and the instrument.
The statistical J- and H-band flux errors are 3.89 \% and 4.51 \%, with systematic errors of 2.98 \% and 3.82 \%.
We conclude that the achievable uncertainties on the absolute photometric amplitudes of
these spectra are not competitive with other measurements
(e.g. the existing 2MASS J and H-band flux errors are 1.57 \% and 2.36 \%, respectively).

The slit mask apodization correction ultimately limits the accuracy of
our absolute calibration measurement and can lead to subtle biases.
However, by connecting them with precise spectral measurements, we can improve
the accuracy of LRS stellar spectra.
The European Space Agency's Gaia \citep{perryman01,jordi10}
mission is a scanning all-sky survey that uses a blue photometer (0.33$\micron$
$<$ $\lambda$ $<$ 0.68$\micron$) and a red (0.64$\micron$ $<$
$\lambda$ $<$ 1.05$\micron$) one to cover 0.33$\micron$ to
1.05$\micron$ with spectral resolution similar to that of the LRS. Because
the Gaia photometers spectrally overlap with the LRS, we expect to eventually be
able to unambiguously correct for the slit mask apodization and
achieve an absolute flux calibration with less than 2 \% accuracy over the
full range $0.4 \leq \lambda \leq 1.6 \, \mu$m for our 105 stars.

In addition, the data reduction procedure described here may be a useful guide for the
Gaia analysis. Since Gaia uses a prism-based photometer source detection,
the data will show a nonlinear spatial variation of constant-wavelength bands
and flux losses by a finite window size, as in our measurements. The background
estimation will also require careful treatment with precise estimation
of the end-to-end Gaia PSF.

\acknowledgments
This work was supported by
NASA APRA research grants NNX07AI54G, NNG05WC18G,
NNX07AG43G, NNX07AJ24G, and NNX10AE12G. Initial support was
provided by an award to J.B. from the Jet Propulsion Laboratory’s
Director’s Research and Development Fund. Japanese participation in
CIBER was supported by KAKENHI (20·34, 18204018, 19540250, 21340047, 21111004, and 26800112)
from Japan Society for the Promotion of
Science (JSPS) and the Ministry of Education, Culture, Sports,
Science, and Technology. Korean participation in CIBER was
supported by the Pioneer Project from the Korea Astronomy and Space
Science Institute.
M.G.K. acknowledges support from the Global PhD Fellowship Program through
the NRF, funded by the Ministry of Education (2011-0007760).
H.M.L. and M.G.L. were supported by NRF grant 2012R1A4A1028713.
M.Z. and P.K. acknowledge support from
NASA postdoctoral program fellowships, and A.C. acknowledges support
from NSF CAREER awards AST-0645427 and NSF AST-1313319.
We thank the dedicated efforts of the sounding rocket staff at the NASA Wallops Flight Facility and White Sands Missile Range
and also thank Dr. Allan Smith, Dr. Keith Lykke, and Dr. Steven Brown (NIST) for the laboratory calibration of the LRS.
This publication makes use of data products from the 2MASS, which is a joint project of the University of Massachusetts
and the Infrared Processing and Analysis Center/California Institute of Technology,
funded by the NASA and the NSF.
This research has made use of the SIMBAD database, operated at CDS, Strasbourg, France,
and the SpeX library .

\clearpage

\clearpage

\clearpage

\clearpage

\clearpage

\begin{figure*}[p]
	\centering
	\epsscale{0.8}
	\plotone{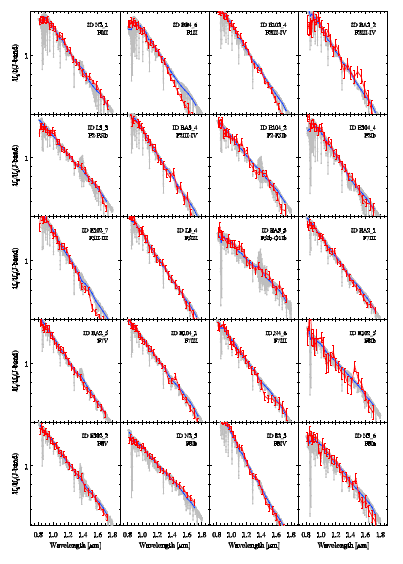}
	\caption{LRS spectra of stars identified in this survey. The blue curve represents the IRTF template degraded
			to fit the observed LRS spectrum, indicated by a red	curve.
			All spectra are normalized at the J-band.
			The original template (gray color) is superimposed
			for comparison. The LRS ID and best-fit IRTF type are indicated on the upper right at each panel. 
			(b)-(f) LRS spectra identified in this work. The color code is the same as that in Figure \ref{sp_fig1}.\label{sp_fig1}}
\end{figure*}

\renewcommand{\thefigure}{\arabic{figure} (Contnued)}
\addtocounter{figure}{-1}

\begin{figure*}[p]
	\centering
	\epsscale{0.8}
	\plotone{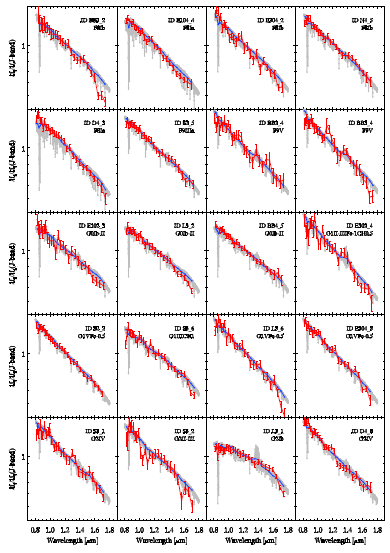}
	\caption{Continued.\label{sp_fig2}}
\end{figure*}

\renewcommand{\thefigure}{\arabic{figure}}

\renewcommand{\thefigure}{\arabic{figure} (Contnued)}
\addtocounter{figure}{-1}

\begin{figure*}[p]
	\centering
	\epsscale{0.8}
	\plotone{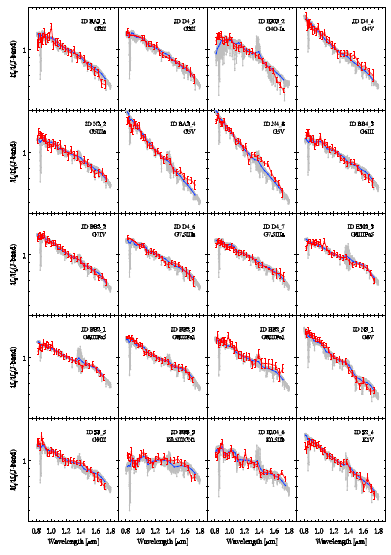}
	\caption{Continued.\label{sp_fig3}}
\end{figure*}

\renewcommand{\thefigure}{\arabic{figure}}

\renewcommand{\thefigure}{\arabic{figure} (Contnued)}
\addtocounter{figure}{-1}

\begin{figure*}[p]
	\centering
	\epsscale{0.8}
	\plotone{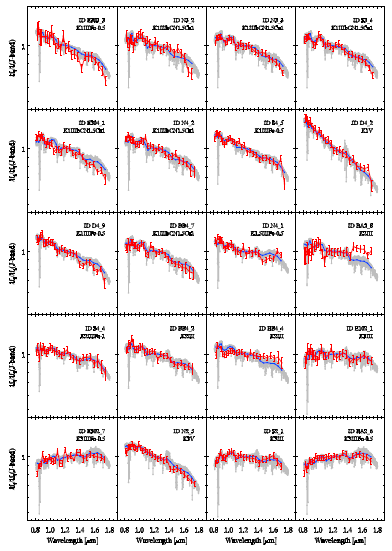}
	\caption{Continued.\label{sp_fig4}}
\end{figure*}

\renewcommand{\thefigure}{\arabic{figure}}

\renewcommand{\thefigure}{\arabic{figure} (Contnued)}
\addtocounter{figure}{-1}

\begin{figure*}[p]
	\centering
	\epsscale{0.8}
	\plotone{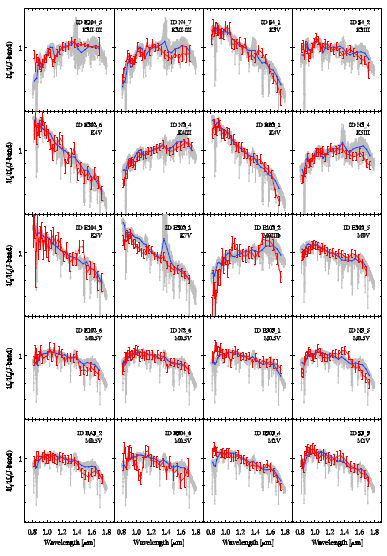}
	\caption{Continued.\label{sp_fig5}}
\end{figure*}

\renewcommand{\thefigure}{\arabic{figure}}

\renewcommand{\thefigure}{\arabic{figure} (Contnued)}
\addtocounter{figure}{-1}

\begin{figure*}[p]
	\centering
	\epsscale{0.8}
	\plotone{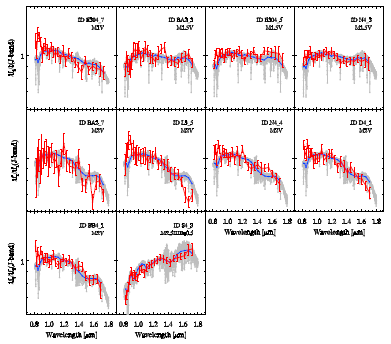}
	\caption{Continued.\label{sp_fig6}}
\end{figure*}

\renewcommand{\thefigure}{\arabic{figure}}

\begin{figure*}[p]
	\centering
	\epsscale{1}
	\plotone{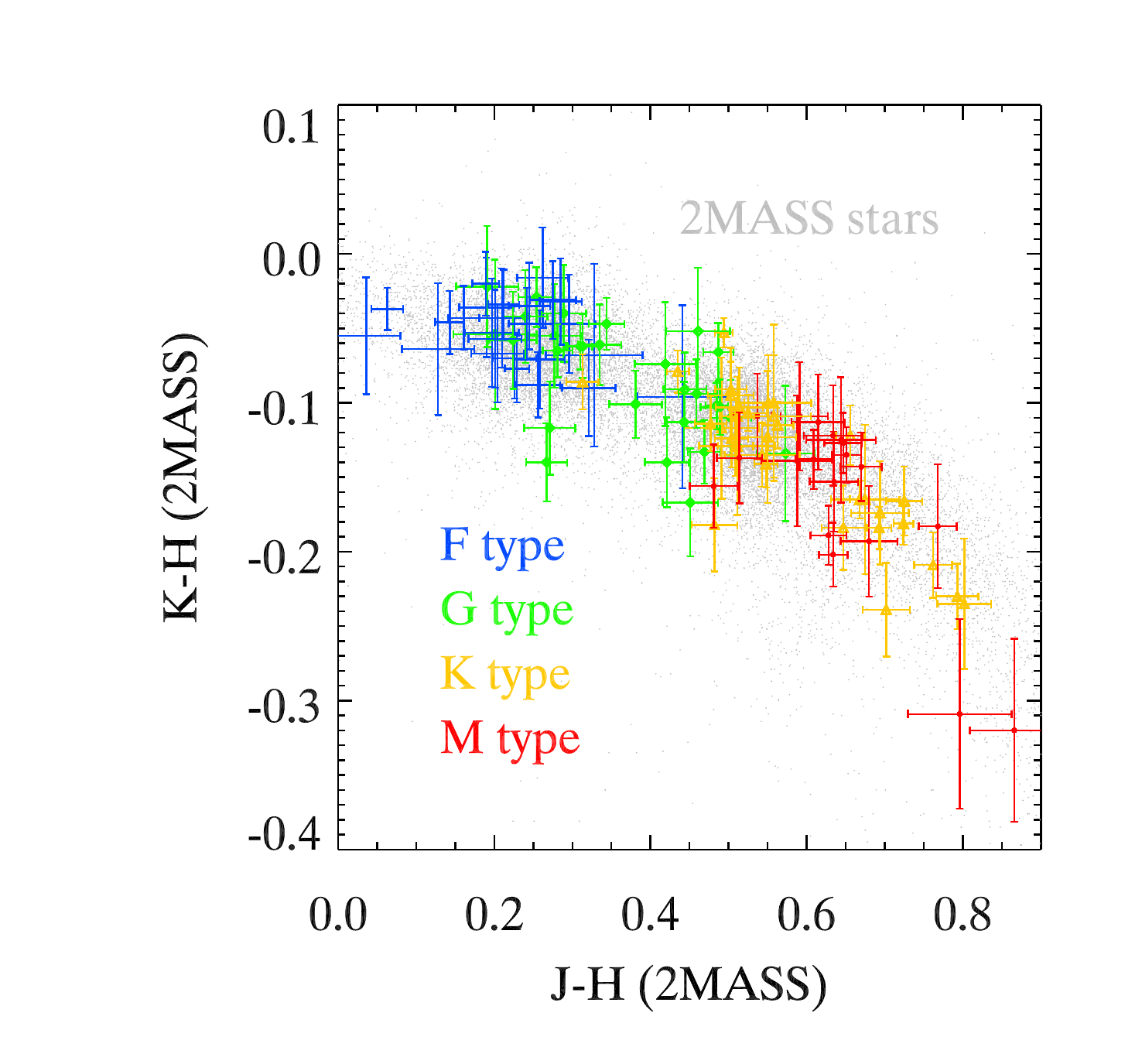}
	\caption{Color-color diagram for all identified
			stars. The J-H and K-H color information is from 2MASS, and the
			type information is from the IRTF fit. The background gray dots
			indicate stars drawn from the 2MASS catalog of each CIBER field.
			The colors represent different stellar types. The scatter of types over the J-H color can be explained either by the noncontinuous IRTF library or by uncertainties in spectral subclass.\label{ccd}}
	
\end{figure*}

\clearpage

\begin{figure*}[p]
	\centering	
	\epsscale{1}
	\plotone{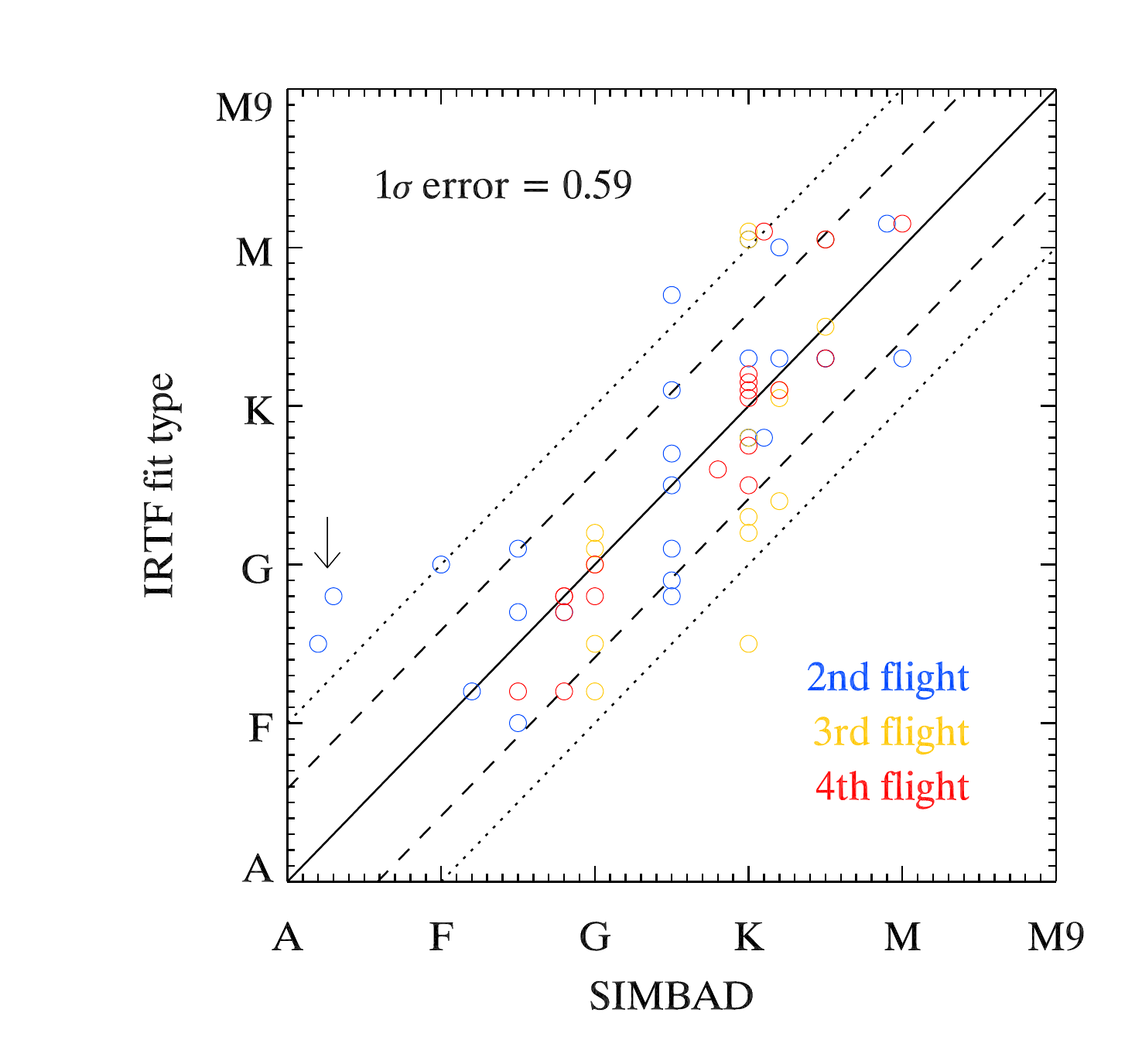}
	\caption{Type comparison determined from the IRTF fit and the literature for 63 stars whose types are already known. The dashed and dotted lines represent the 1$\sigma$ error and $\pm$1 spectral type, respectively. The colors represent the different flights' data. Two \textit{A}-type stars, indicated by an arrow, are fitted to \textit{F}-type stars. Fit types based on the Pickles library also give the same results.\label{type_difference}}
\end{figure*}

\clearpage

\begin{figure*}[p]
	\centering
	\epsscale{0.8}
	\plotone{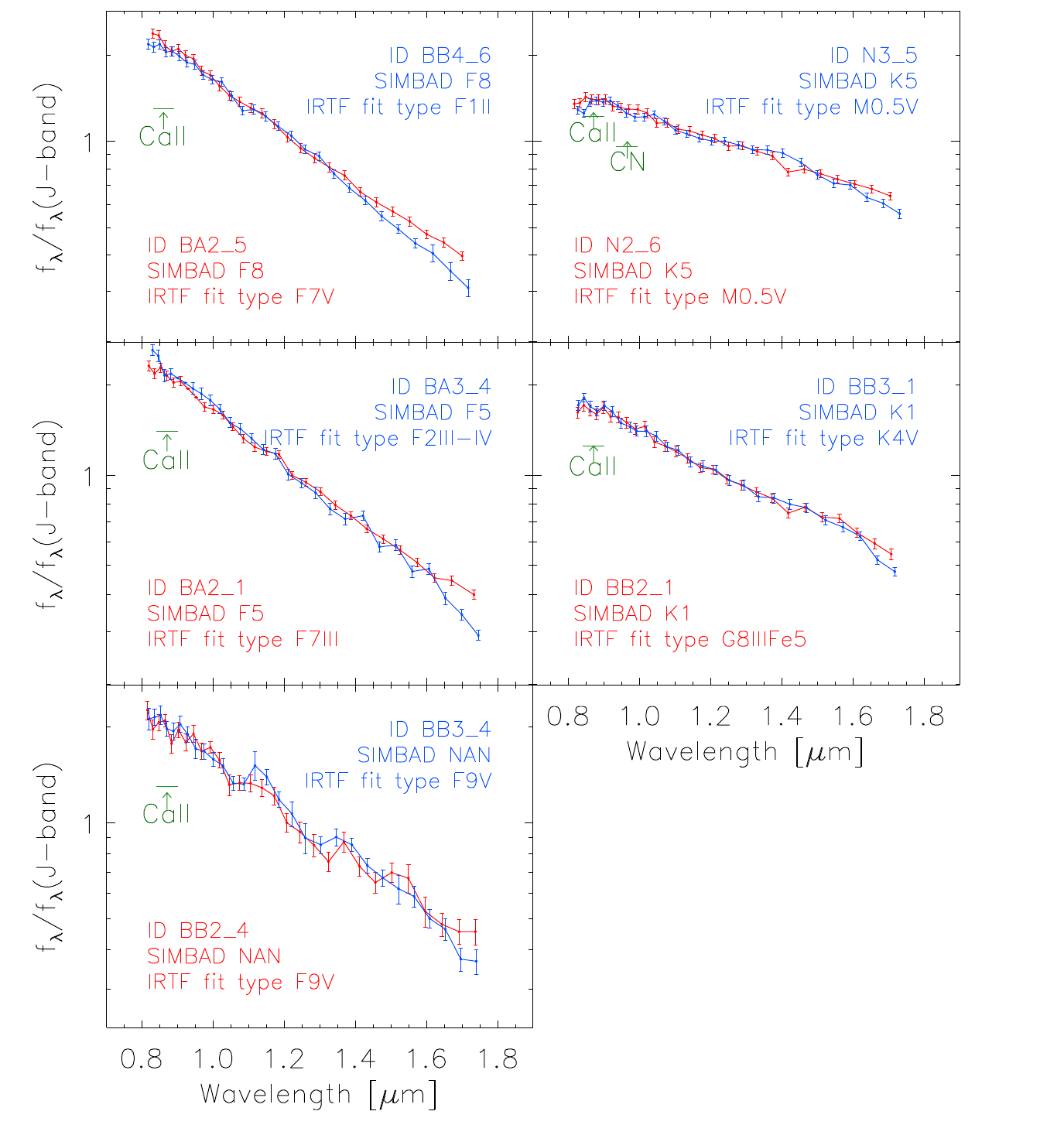}
	\caption{Five stars are serendipitously observed in two independent flights. Each panel shows two spectra extracted from each flight. Top left panel: 2nd flight (BA2\_5), 4th flight (BB4\_6). Top right panel: 2nd flight (N2\_6), 3rd flight (N3\_5). Middle left panel: 2nd flight (BA2\_1), 3rd flight (BA3\_4). Middle right panel: 2nd flight (BB2\_1), 3rd flight (BB3\_1). Bottom left panel: 2nd flight (BB2\_4), 3rd flight (BB3\_4). The large discrepancies arise from calibration error above 1.6 $\micron$ but show consistency of in-flight calibration below 1.6 $\micron$. \label{dup_star}}
\end{figure*}

\clearpage

\begin{figure*}[p]
	\centering
	\epsscale{0.7}
	\plotone{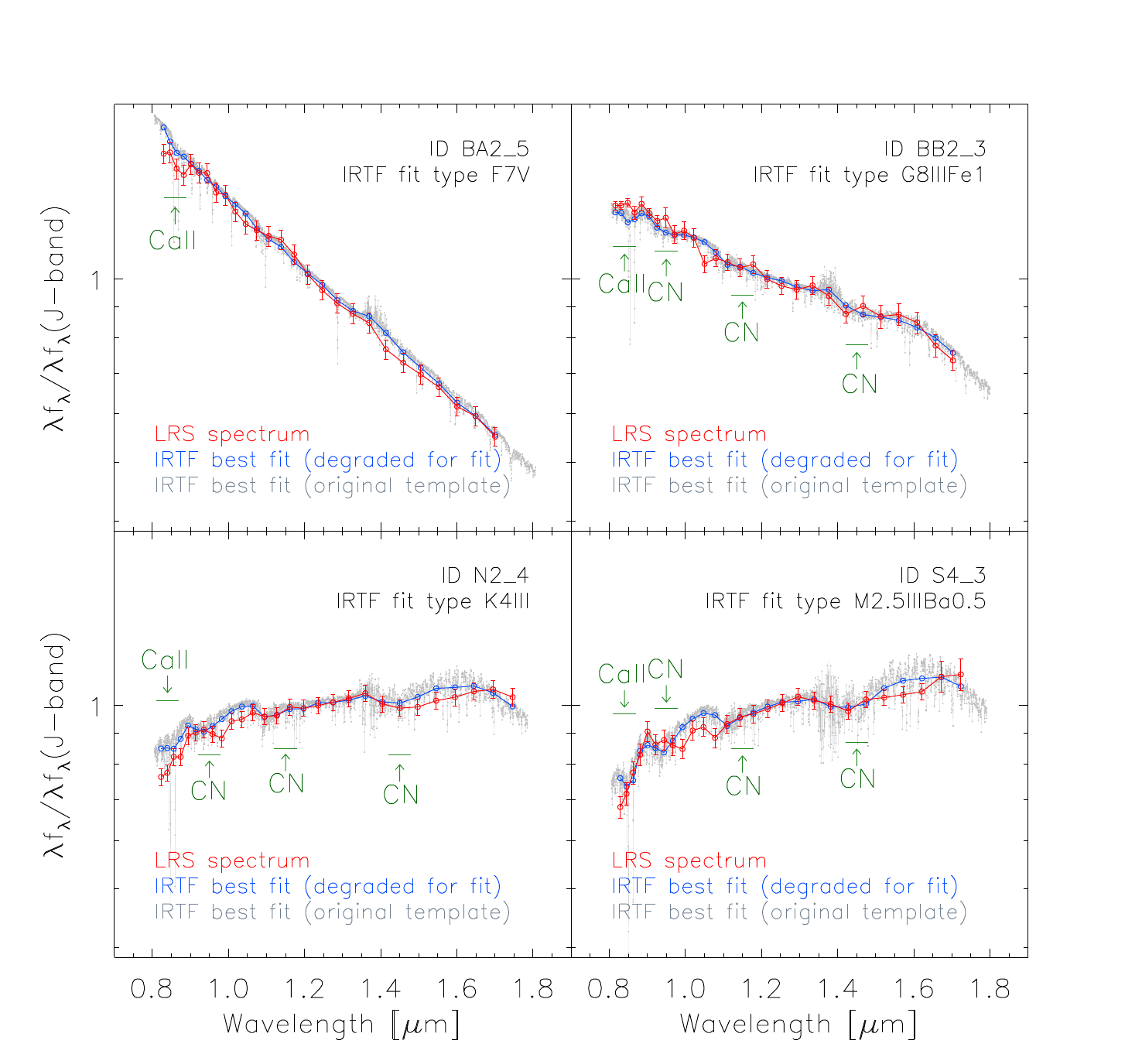}
	\caption{Representative examples of LRS spectra from this work. 
		The color code is the same as that in Figure \ref{sp_fig1}.
     	 F, G, K, and M stellar types are shown in each
		panel. Compared to other types, a typical F-type spectrum
		(top left panel) does not show any obvious absorption
		features across the wavelength range. We identified several
		features in our LRS spectra that correspond to
		typical absorption lines in the near-IR (i.e., CaII with
		bandhead at 0.85 $\micron$, CN with bandhead at 0.95, 1.15, and 1.5 $\micron$). The strongest feature in the F-type stars (top left) is the CaII triplet line, indicated with an arrow at 0.85 $\micron$. From types later than G (top right), CN bands appear with bandheads at 1.1, 0.91, 0.94, and 1.4$\micron$. We also identified M-type stars, as indicated in the bottom right panel. Since M-type stars have dominant molecular bands in their spectra, the identified lines are blended with other strong molecular bands, such as TiO (bandhead at 0.82$\micron$), ZrO (bandhead at 0.93$\micron$), FeH (bandhead at 0.99$\micron$), and H${_2}$O (bandhead at 1.4$\micron$). The strength of each line depends on the spectral type.\label{type_example}}
\end{figure*}

\clearpage

\begin{figure*}[p]
	\centering
	\epsscale{1}
	\plotone{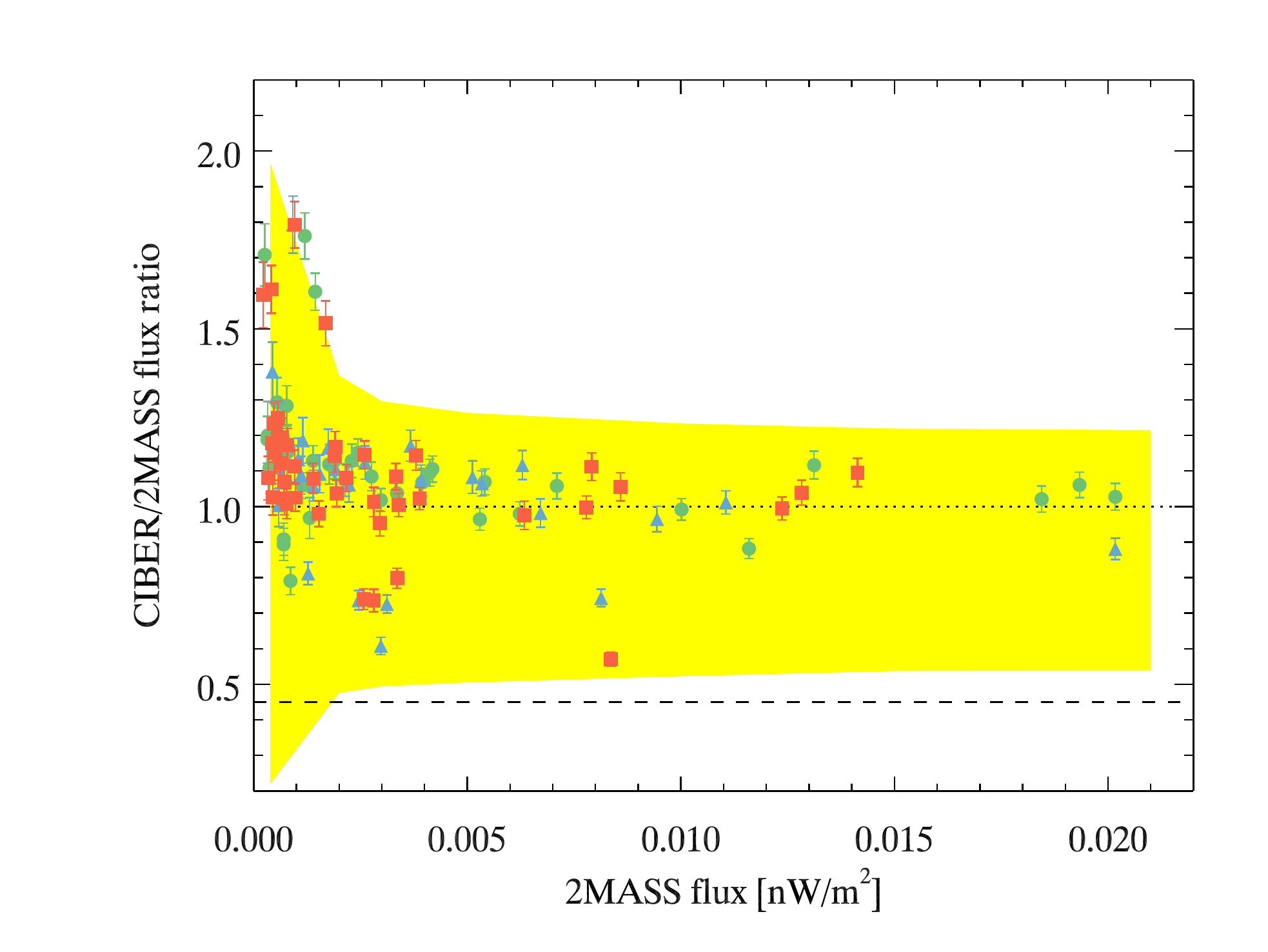}
	\caption{Flux ratios of all LRS stars to the matched 2MASS stars in the J-band.
     		Each color represents the stars observed from each flight.
	     	Since the LRS flux is apodized by the slit mask, an aperture correction has been made to
		    yield ratio unity in the ideal case (dotted line). The averaged original flux ratio is drawn as a dashed line, and its reciprocal is used for aperture correction. The color-shaded area shows the range of relation we expect from an instrument simulation, representing the upper and lower bounds of the absolute calibrations of the LRS.\label{model_J}}
\end{figure*}

\clearpage

\begin{figure*}[p]
	\centering
	\epsscale{1}
	\plotone{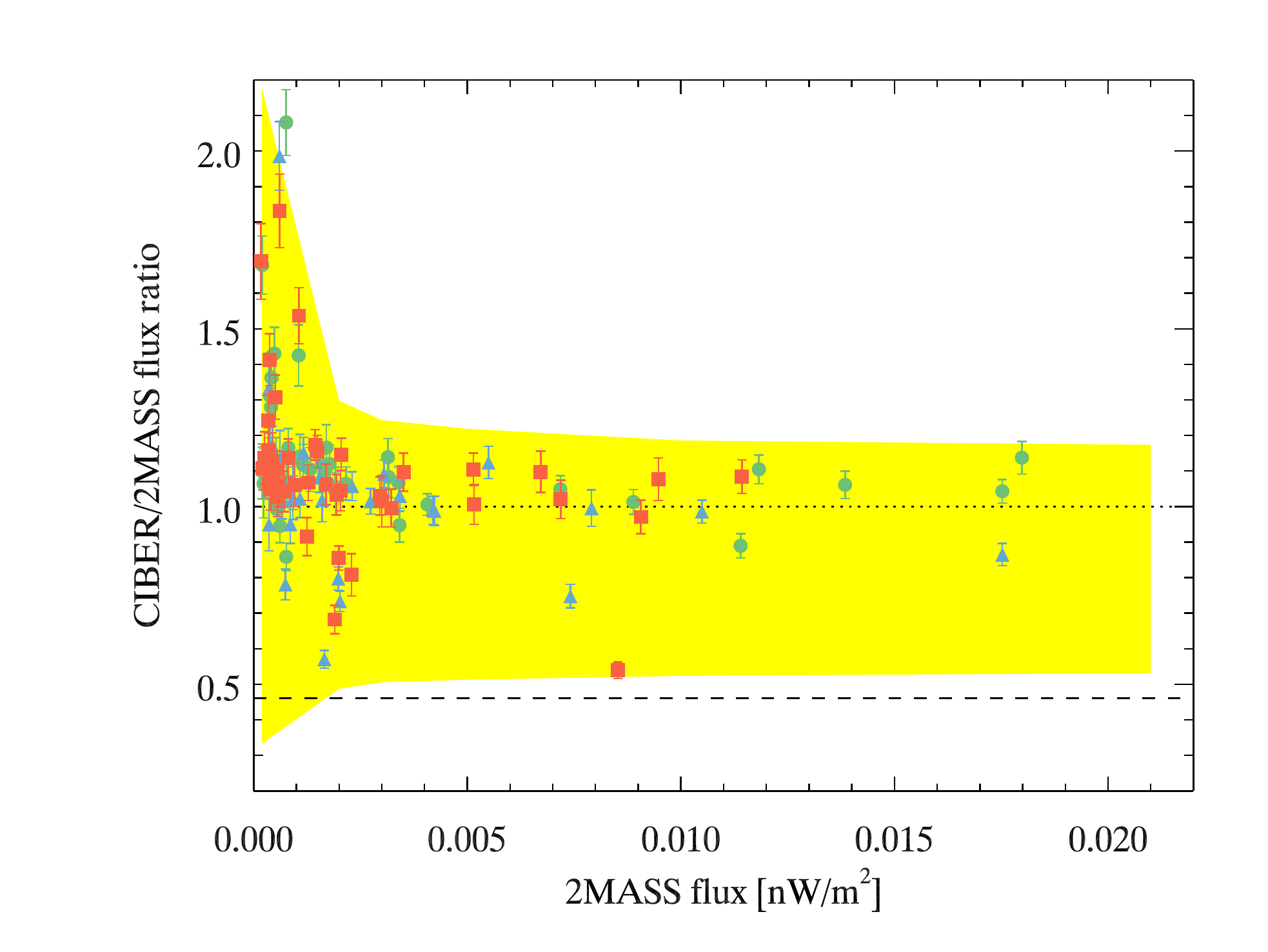}
	\caption{Same as Figure \ref{model_J} but for the H-band.\label{model_H}}
\end{figure*}

\clearpage

\begin{deluxetable}{ccc}
	\tabletypesize{\scriptsize}
	\tablecaption{ Rocket-Commanded Coordinates for the observed field. Arabic numbers after the Hyphen for the Elat fields indicate the flight number \label{tbl1}}
	\tablewidth{0pt}
	\tablehead{
		\colhead{Field} & \colhead{R.A.} & \colhead{Decl.}
	}	
	\startdata
	Elat10-2 & 15:07:60.0 & -2:00:00 \\
	Elat30-2 & 14:44:00 & 20:00:00 \\
	Elat30-3 & 15:48:00 & 9:30:00 \\
	Elat10-4 & 12:44:00 & 8:00:00 \\
	Elat30-4 & 12:52:00 & 27:00:00 \\
	NEP & 18:00:00 & 66:20:23.987 \\
	SWIRE & 16:11:00 & 55:00:00 \\
	BootesA & 14:33:54.719 & 34:53:2.396 \\
	BootesB & 14:29:17.761 & 34:53:2.396 \\
	Lockman & 10:45:12.0 & 58:00:00 \\
	DGL & 16:47:60.0 & 69:00:00
	\enddata
\end{deluxetable}

\clearpage

\begin{deluxetable}{ccccccc}
	\tabletypesize{\scriptsize}
	\tablecaption{ Number of stars rejected at each criterion \label{tbl3}}
	\tablewidth{0pt}
	\tablehead{
		\colhead{Flight} & \colhead{Total Candidates} & \colhead{Crit.(\textit{i})} & \colhead{Crit.(\textit{ii})} & \colhead{Crit.(\textit{iii})} & \colhead{Crit.(\textit{iv})} & \colhead{Total in Final Catalog}
	}	
	\startdata
	2nd flight & 198 & 15 & 43 & 8 & 145 & 38 \\
	3rd flight & 177 & 14 & 41 & 6 & 127 & 30 \\
	4th flight & 171 & 23 & 43 & 5 & 117 & 42
	\enddata
\end{deluxetable}

\clearpage

\begin{landscape}
	\begin{deluxetable}{cccccccccccccc}
	\tabletypesize{\tiny}
	\setlength{\tabcolsep}{0.02in}
	\tablewidth{0pt}
	\tablecaption{ Star catalog \label{tbl2}}
	\tablehead{
	\colhead{Flight} & \colhead{Field} & \colhead{ID} & \colhead{Name} & \colhead{R.A.\tablenotemark{a}} &
		\colhead{Decl.\tablenotemark{b}} & \colhead{LRS J\tablenotemark{c}} & \colhead{LRS H\tablenotemark{d}} &
		\colhead{2MASS J\tablenotemark{e}} & \colhead{2MASS H\tablenotemark{f}} &
		\colhead{SIMBAD type\tablenotemark{g}} & \colhead{Best-fit IRTF Type} & \colhead{$\chi^2$} & \colhead{Note}
	}\startdata
&   Elat10  &  E102\_1  &                   TYC5000-614-1  &      15:06:50.134  &      -00:02:47.746  &       9.020  &       8.283  &       8.283  &       7.608  &       K2  &                           K3III  &       0.720  &                             ...  \\
&   Elat10  &  E102\_2  &                             ...  &      14:59:05.568  &      -01:08:23.294  &       9.095  &       8.279  &       8.350  &       7.484  &      ...  &                          M0IIIb  &       4.582  &                             ...  \\
&   Elat10  &  E102\_3  &                        HD131553  &     14:54:20.898  &     -01:52:19.938  &       9.576  &       9.241  &       8.673  &       8.472  &      F0V  &                         G0Ib-II  &       0.522  &                             ...  \\
&   Elat10  &  E102\_4  &                        HD134456  &      15:09:58.320  &     -00:52:47.269  &       7.872  &       7.754  &       6.982  &       6.854  &    F2III  &                        F2III-IV  &       0.076  &                             ...  \\
&   Elat10  &  E102\_5  &                   TYC5001-847-1  &     15:14:43.328  &     -01:31:43.763  &       9.940  &       9.633  &       9.226  &       8.898  &      ...  &                            F8Ib  &       0.416  &                             ...  \\
&   Elat10  &  E102\_6  &                      BD-01-3038  &     15:14:15.481  &      -01:37:09.268  &       8.273  &       7.633  &       7.477  &       6.862  &       K0  &                           M0.5V  &       0.462  &                             ...  \\
&   Elat10  &  E102\_7  &                        HD133213  &      15:03:28.468  &      -03:10:05.732  &       8.802  &       8.751  &       8.066  &       8.030  &    A2III  &                        F5II-III  &       0.086  &                             ...  \\
&   Elat30  &  E302\_1  &                      BD+22-2745  &      14:46:03.405  &      22:04:37.528  &       8.065  &       7.499  &       7.158  &       6.664  &       G5  &                             K7V  &       0.304  &                             ...  \\
&   Elat30  &  E302\_2  &                        HD127666  &      14:32:02.149  &      22:04:47.600  &       8.645  &       8.396  &       7.866  &       7.676  &       G5  &                             F8V  &       0.045  &                             ...  \\
&   Elat30  &  E302\_3  &                        HD131132  &     14:51:16.019  &     18:38:59.284  &       6.648  &       6.111  &       5.803  &       5.334  &       K0  &                        G8IIIFe5  &       0.260  &                             ...  \\
&   Elat30  &  E302\_4  &                      BD+19-2867  &     14:49:56.793  &     18:37:29.741  &      10.875  &      10.668  &      10.195  &       9.928  &       G5  &               G1II-IIIFe-1CH0.5  &       0.705  &                             ...  \\
&   Elat30  &  E302\_5  &                      BD+19-2857  &     14:45:32.922  &     18:40:20.255  &       7.342  &       6.643  &       6.466  &       5.815  &       K2  &                             M0V  &       0.234  &                             ...  \\
&   Elat30  &  E302\_6  &                   TYC1481-620-1  &     14:46:48.921  &     17:30:12.359  &      10.208  &       9.644  &       9.620  &       9.138  &      ...  &                             K4V  &       0.551  &                             ...  \\
&   Elat30  &  E302\_7  &                      BD+18-2928  &     14:45:45.544  &     17:30:17.950  &       6.555  &       5.752  &       5.752  &       5.050  &       M0  &                     K3IIIFe-0.5  &       1.488  &                             ...  \\
\multirow{3}{*}{2nd} &      NEP  &    N2\_1  &                       BD+68-954  &     17:43:43.944  &     68:24:26.593  &      10.067  &       9.742  &       9.394  &       9.168  &       F5  &                            F0II  &       0.064  &                             ...  \\
&      NEP  &    N2\_2  &                             ...  &     17:38:56.867  &     66:22:12.587  &      10.726  &      10.216  &      10.440  &       9.937  &      ...  &                          G5IIIa  &       0.240  &                             ...  \\
&      NEP  &    N2\_3  &                     BD+67-1039A  &     17:52:45.953  &      67:00:12.935  &       8.925  &       8.587  &       8.571  &       8.130  &      ...  &                            F8Ib  &       0.045  &                             ...  \\
&      NEP  &    N2\_4  &                   TYC4208-116-1  &     17:49:23.407  &     65:28:22.606  &       7.646  &       6.837  &       6.840  &       6.047  &      ...  &                           K4III  &       0.807  &                             ...  \\
&      NEP  &    N2\_5  &                      BD+67-1067  &     18:20:50.229  &      67:55:01.776  &       8.199  &       7.694  &       7.430  &       6.939  &       K0  &                             K3V  &       0.119  &                             ...  \\
&      NEP  &    N2\_6\tablenotemark{i}  &                        HD166779  &      18:07:35.504  &     63:54:12.298  &       6.544  &       5.874  &       5.706  &       5.078  &       K5  &                           M0.5V  &       0.221  &                             ...  \\
&    SWIRE  &    S2\_1  &                        HD144245  &      16:01:58.920  &      56:36:03.496  &       6.921  &       6.238  &       6.173  &       5.505  &       K5  &                           K3III  &       0.201  &                             ...  \\
&    SWIRE  &    S2\_2  &                        HD144082  &       16:01:09.819  &     56:26:23.172  &       7.929  &       7.644  &       7.135  &       6.944  &       F5  &                       G1VFe-0.5  &       0.051  &                             ...  \\
&    SWIRE  &    S2\_3  &                        HD147733  &     16:20:51.242  &     54:23:10.320  &       8.172  &       8.125  &       7.414  &       7.351  &       A3  &                            F8IV  &       0.059  &                             ...  \\
&    SWIRE  &    S2\_4  &                        HD234317  &     16:32:27.630  &     54:20:14.320  &       8.713  &       8.283  &       7.999  &       7.564  &       G5  &                             K1V  &       0.081  &                             ...  \\
&    SWIRE  &    S2\_5  &                        HD146736  &     16:15:15.896  &      52:01:48.338  &       8.929  &       8.618  &       8.140  &       7.884  &       G5  &                          F9IIIa  &       0.060  &                             ...  \\
&  BootesA  &   BA2\_1\tablenotemark{j}  &                        HD126878  &     14:27:13.534  &     34:43:19.996  &       8.631  &       8.385  &       7.783  &       7.640  &       F5  &                           F7III  &       0.046  &                             ...  \\
&  BootesA  &   BA2\_2  &                   TYC2557-719-1  &     14:41:46.727  &     33:34:23.452  &      10.800  &      10.557  &      10.045  &       9.783  &      ...  &                        F2III-IV  &       0.331  &                             ...  \\
&  BootesA  &   BA2\_3  &                   TYC2556-652-1  &     14:33:46.073  &     33:34:53.886  &      10.341  &       9.620  &       9.352  &       8.717  &      K9V  &                           M1.5V  &       1.125  &              high-proper-motion  \\
&  BootesA  &   BA2\_4  &                      BD+34-2527  &     14:25:57.827  &     33:34:32.984  &       9.846  &       9.426  &       9.250  &       8.973  &    G5III  &                             G5V  &       0.120  &                             ...  \\
&  BootesA  &   BA2\_5\tablenotemark{h}  &                        HD126210  &     14:23:24.060  &     33:34:19.099  &       8.480  &       8.274  &       7.653  &       7.492  &       F8  &                             F7V  &       0.039  &                             ...  \\
&  BootesA  &   BA2\_6  &                      BD+34-2522  &     14:21:54.490  &     33:34:35.580  &       7.311  &       6.514  &       6.307  &       5.545  &       K5  &                     K3IIIFe-0.5  &       0.584  &                             ...  \\
&  BootesA  &   BA2\_7  &                             ...  &     14:41:50.085  &     32:24:33.790  &      10.848  &      10.330  &      10.178  &       9.587  &      ...  &                             M2V  &       1.521  &                             ...  \\
&  BootesA  &   BA2\_8  &                   TYC2553-127-1  &     14:29:10.917  &     32:27:40.871  &      10.252  &       9.490  &       9.130  &       8.483  &      ...  &                           K2III  &       1.255  &                             ...  \\
&  BootesB  &   BB2\_1\tablenotemark{k}  &                  TYC2560-1157-1  &     14:38:39.909  &     35:31:13.224  &       9.347  &       8.799  &       8.611  &       8.100  &       K1  &                        G8IIIFe5  &       0.143  &                             ...  \\
&  BootesB  &   BB2\_2  &                      BD+36-2489  &     14:24:52.634  &     35:32:12.714  &       9.026  &       8.530  &       8.773  &       8.484  &       G5  &                            G7IV  &       0.107  &                             ...  \\
&  BootesB  &   BB2\_3  &                      BD+32-2490  &      14:34:03.366  &       32:06:02.588  &       9.640  &       9.089  &       8.835  &       8.414  &       K0  &                        G8IIIFe1  &       0.127  &                             ...  \\
&  BootesB  &   BB2\_4\tablenotemark{l}  &                      BD+31-2630  &      14:33:01.264  &     30:56:33.554  &      10.240  &       9.793  &       9.504  &       9.246  &      ...  &                             F9V  &       0.336  &                             ...  \\
&  BootesB  &   BB2\_5  &                   TYC2553-961-1  &     14:24:21.497  &      30:58:03.684  &      10.323  &       9.713  &       9.351  &       8.864  &      ...  &                        G8IIIFe1  &       0.580  &                             ...  \\\hline
&   Elat30  &  E303\_1  &                      BD+11-2874  &      15:52:08.230  &     10:52:28.103  &       7.882  &       7.169  &       6.692  &       6.012  &      K5V  &                           M0.5V  &       0.330  &            spectroscopic binary  \\
&   Elat30  &  E303\_2  &                        HD141631  &     15:49:47.057  &     10:48:24.520  &       8.251  &       7.922  &       7.555  &       7.096  &       K2  &                          G4O-Ia  &       0.206  &                             ...  \\
&   Elat30  &  E303\_3  &                    TYC947-300-1  &     15:50:53.577  &      09:41:15.828  &      10.379  &       9.841  &       9.861  &       9.310  &      ...  &                     K1IIIFe-0.5  &       0.595  &                             ...  \\
&   Elat30  &  E303\_4  &                        HD141531  &     15:49:16.496  &      09:36:42.408  &       7.718  &       7.052  &       6.971  &       6.337  &        K  &                             M1V  &       0.089  &                             ...  \\
&      NEP  &    N3\_1  &                        HD164781  &      17:57:03.647  &     68:49:19.744  &       8.948  &       8.601  &       7.733  &       7.423  &       K0  &                             G8V  &       0.076  &                             ...  \\
&      NEP  &    N3\_2  &                  TYC4428-1122-1  &     17:54:46.231  &      68:06:42.016  &       9.753  &       9.250  &       9.009  &       8.353  &      ...  &                  K1IIIbCN1.5Ca1  &       0.629  &                             ...  \\
&      NEP  &    N3\_3  &                      BD+67-1050  &      18:06:45.898  &     67:50:40.686  &       8.273  &       7.722  &       7.485  &       6.976  &       K2  &                  K1IIIbCN1.5Ca1  &       0.134  &                             ...  \\
&      NEP  &    N3\_4  &                      BD+65-1248  &     18:12:21.398  &     65:36:17.381  &       7.214  &       6.492  &       6.359  &       5.635  &       K5  &                           K5III  &       0.919  &                             ...  \\
&      NEP  &    N3\_5\tablenotemark{i}  &                        HD166779  &      18:07:35.504  &     63:54:12.298  &       6.711  &       6.077  &       5.706  &       5.078  &       K5  &                           M0.5V  &       0.455  &                             ...  \\
&      NEP  &    N3\_6  &                   TYC4226-812-1  &     18:25:26.020  &      66:00:38.783  &       9.655  &       9.417  &       8.924  &       8.714  &      ...  &                            F8Ia  &       0.293  &                             ...  \\
&    SWIRE  &    S3\_1  &                      BD+55-1802  &      16:01:45.359  &     54:48:40.882  &      10.325  &      10.033  &       9.570  &       9.330  &       G0  &                            G2IV  &       0.392  &                             ...  \\
&    SWIRE  &    S3\_2  &                  TYC3870-1085-1  &     15:54:21.929  &     53:36:47.786  &      10.417  &      10.198  &       9.554  &       9.300  &      ...  &                        G2II-III  &       0.871  &                             ...  \\
\multirow{3}{*}{3rd} &    SWIRE  &    S3\_3  &                   TYC3870-366-1  &     15:53:29.099  &     53:28:36.008  &       8.669  &       8.062  &       7.928  &       7.281  &      ...  &                             M1V  &       0.285  &                             ...  \\
&    SWIRE  &    S3\_4  &                   TYC3877-704-1  &     16:10:22.667  &     54:28:38.784  &       9.017  &       8.472  &       8.258  &       7.715  &      ...  &                  K1IIIbCN1.5Ca1  &       0.239  &                             ...  \\
&    SWIRE  &    S3\_5  &                  TYC3877-1592-1  &      16:01:43.031  &      53:06:25.855  &      10.233  &       9.746  &       9.566  &       9.077  &      ...  &                           G9III  &       0.136  &                             ...  \\
&    SWIRE  &    S3\_6  &                   TYC3878-216-1  &     16:25:31.829  &     53:25:25.453  &       9.065  &       8.709  &       8.364  &       8.020  &      ...  &                        G1IIICH1  &       0.214  &                             ...  \\
&  Lockman  &    L3\_1  &                        V*DM-UMa  &     10:55:43.521  &      60:28:09.613  &       7.975  &       7.476  &       7.194  &       6.621  &    K0III  &                            G2Ib  &       0.233  &                             ...  \\
&  Lockman  &    L3\_2  &                         HD94880  &     10:58:21.518  &     59:16:53.422  &       7.787  &       7.482  &       6.900  &       6.629  &       G0  &                         G0Ib-II  &       0.115  &                             ...  \\
&  Lockman  &    L3\_3  &                         HD92320  &     10:40:56.905  &     59:20:33.065  &       7.947  &       7.662  &       7.148  &       6.852  &       G0  &                         F2-F5Ib  &       0.109  &              high-proper-motion  \\
&  Lockman  &    L3\_4  &                        HD237955  &     10:57:44.114  &      58:10:01.103  &       9.799  &       9.619  &       8.705  &       8.508  &       G0  &                           F5III  &       0.038  &                             ...  \\
&  Lockman  &    L3\_5  &                   TYC3827-847-1  &      11:01:59.570  &     56:58:11.510  &       9.498  &       9.094  &       8.816  &       8.279  &      ...  &                             M2V  &       0.479  &                             ...  \\
&  Lockman  &    L3\_6  &                        HD237961  &      11:00:12.007  &     56:59:49.481  &       9.267  &       9.049  &       8.495  &       8.271  &       G0  &                       G1VFe-0.5  &       0.304  &                             ...  \\
&  BootesA  &   BA3\_1  &                       BD+362491  &      14:26:05.241  &      35:50:00.776  &       8.897  &       8.498  &       8.095  &       7.676  &       K0  &                            G3II  &       0.515  &                             ...  \\
&  BootesA  &   BA3\_2  &                        HD128368  &     14:35:32.053  &     34:41:11.540  &       7.436  &       6.789  &       6.530  &       5.942  &       K0  &                           M0.5V  &       0.215  &                             ...  \\
&  BootesA  &   BA3\_3  &                      BD+35-2576  &     14:32:31.567  &      34:42:09.493  &       9.291  &       8.834  &       9.058  &       8.737  &       K0  &                       F5Ib-G1Ib  &       0.143  &                             ...  \\
&  BootesA  &   BA3\_4\tablenotemark{j}  &                        HD126878  &     14:27:13.534  &     34:43:19.996  &       9.190  &       9.091  &       7.783  &       7.640  &       F5  &                        F2III-IV  &       0.060  &                             ...  \\
&  BootesB  &   BB3\_1\tablenotemark{k}  &                  TYC2560-1157-1  &     14:38:39.909  &     35:31:13.224  &       9.416  &       8.918  &       8.611  &       8.100  &       K1  &                             K4V  &       0.124  &                             ...  \\
&  BootesB  &   BB3\_2  &                      BD+32-2503  &      14:41:07.455  &      32:04:45.095  &       9.628  &       9.449  &       8.853  &       8.624  &      ...  &                            F8Ib  &       0.198  &                             ...  \\
&  BootesB  &   BB3\_3  &                      BD+32-2456  &     14:18:52.718  &      32:06:31.003  &       9.191  &       8.531  &       7.992  &       7.444  &    K2III  &                      K0.5IIICN1  &       0.534  &                             ...  \\
&  BootesB  &   BB3\_4\tablenotemark{l}  &                      BD+31-2630  &      14:33:01.264  &     30:56:33.554  &      10.170  &       9.940  &       9.504  &       9.246  &      ...  &                             F9V  &       0.438  &                             ...  \\\hline
&   Elat10  &  E104\_1  &                        HD111645  &     12:50:42.449  &      08:52:30.238  &       8.908  &       8.691  &       8.124  &       7.920  &       F8  &                           F7III  &       0.041  &                             ...  \\
&   Elat10  &  E104\_2  &                      BD+11-2491  &      12:46:07.870  &      11:09:25.744  &      10.229  &       9.992  &       9.486  &       9.201  &       F8  &                         F2-F5Ib  &       0.162  &                             ...  \\
&   Elat10  &  E104\_3  &                             ...  &     12:41:28.720  &     10:52:57.907  &      10.959  &      10.368  &      10.599  &      10.096  &      ...  &                             K5V  &       0.702  &                             ...  \\
&   Elat10  &  E104\_4  &                        HD110777  &     12:44:20.102  &      06:51:16.916  &       8.442  &       8.212  &       7.663  &       7.418  &       G0  &                            F8Ia  &       0.148  &                             ...  \\
&   Elat10  &  E104\_5  &                      BD+10-2440  &     12:33:51.920  &      09:31:54.156  &       8.139  &       7.372  &       6.662  &       5.860  &      ...  &                        K3II-III  &       1.012  &                             ...  \\
&   Elat10  &  E104\_6  &                        HD109824  &     12:37:48.044  &       04:59:07.195  &       6.860  &       6.296  &       6.092  &       5.542  &       K0  &                         K0.5IIb  &       0.570  &                             ...  \\
&   Elat30  &  E304\_1  &                             ...  &      13:02:54.144  &     26:23:27.762  &       8.966  &       8.441  &       8.267  &       7.756  &      ...  &                  K1IIIbCN1.5Ca1  &       0.478  &                             ...  \\
&   Elat30  &  E304\_2  &                      BD+27-2207  &      13:02:50.671  &      26:50:00.402  &      10.924  &      10.630  &      10.141  &       9.899  &       F8  &                            F8Ib  &       0.262  &                             ...  \\
&   Elat30  &  E304\_3  &                   TYC1995-264-1  &      13:02:50.439  &     27:29:22.283  &      10.212  &      10.004  &       9.586  &       9.251  &      ...  &                       G1VFe-0.5  &       0.121  &                             ...  \\
&   Elat30  &  E304\_4  &                      BD+27-2197  &     12:57:45.577  &      27:01:51.600  &      10.562  &      10.374  &       9.873  &       9.672  &       F5  &                            F2Ib  &       0.098  &                             ...  \\
&   Elat30  &  E304\_5  &                  TYC1995-1123-1  &     12:57:25.736  &     28:18:25.992  &       9.837  &       9.006  &       8.997  &       8.229  &      ...  &                           M1.5V  &       0.608  &                             ...  \\
&   Elat30  &  E304\_6  &                       LP322-154  &      12:57:04.818  &     29:30:36.860  &      10.454  &       9.808  &       9.740  &       9.096  &      K5V  &                           M0.5V  &       1.460  &              high-proper-motion  \\
&   Elat30  &  E304\_7  &                   TYC2532-820-1  &     12:56:45.236  &     30:44:22.556  &      10.678  &      10.006  &       9.838  &       9.324  &      K1V  &                             M1V  &       0.344  &                             ...  \\
&      NEP  &    N4\_1  &                       BD+68-951  &     17:38:51.760  &     68:13:16.536  &       9.137  &       8.449  &       7.942  &       7.438  &       K0  &                   K1.5IIIFe-0.5  &       0.273  &                   multiple-star  \\
&      NEP  &    N4\_2  &                        HD161500  &     17:41:10.318  &     65:13:10.301  &       7.442  &       6.860  &       6.633  &       6.119  &       K2  &                  K1IIIbCN1.5Ca1  &       0.312  &                             ...  \\
&      NEP  &    N4\_3  &                         G227-20  &     17:52:11.850  &      64:46:08.720  &       9.077  &       8.391  &       8.249  &       7.615  &      M0.5V  &                           M1.5V  &       0.449  &              high-proper-motion  \\
\multirow{3}{*}{4th} &      NEP  &    N4\_4  &                  TYC4208-1599-1  &      17:52:05.421  &     64:37:15.827  &      10.278  &       9.725  &       9.929  &       9.259  &      ...  &                             M2V  &       0.486  &                             ...  \\
&      NEP  &    N4\_5  &                     BD+64-1227A  &     17:52:17.178  &     64:14:16.411  &       8.816  &       8.500  &       8.400  &       8.125  &      ...  &                            F8Ib  &       0.046  &                             ...  \\
&      NEP  &    N4\_6  &                   TYC4213-161-1  &      18:03:24.923  &     67:12:41.681  &      10.171  &       9.868  &       9.327  &       9.115  &      ...  &                           F7III  &       0.109  &                             ...  \\
&      NEP  &    N4\_7  &                      BD+66-1074  &      18:03:15.008  &     66:20:29.069  &       7.609  &       6.866  &       6.739  &       6.046  &       K5  &                        K3II-III  &       1.262  &                             ...  \\
&      NEP  &    N4\_8  &                        HD170592  &     18:25:24.759  &     65:45:34.470  &       7.474  &       7.143  &       6.722  &       6.409  &       K0  &                             G5V  &       0.148  &                             ...  \\
&    SWIRE  &    S4\_1  &                  TYC3870-1026-1  &     15:55:16.319  &     54:45:12.510  &      10.127  &       9.564  &       9.332  &       8.829  &      ...  &                             K3V  &       0.261  &                             ...  \\
&    SWIRE  &    S4\_2  &                  TYC3496-1361-1  &      15:56:04.610  &     52:13:29.543  &       8.240  &       7.566  &       7.519  &       6.825  &      ...  &                           K3III  &       0.421  &                             ...  \\
&    SWIRE  &    S4\_3  &                  TYC3880-1133-1  &      16:03:15.627  &      56:02:35.210  &       8.711  &       7.821  &       7.791  &       6.995  &      ...  &                    M2.5IIIBa0.5  &       2.347  &                             ...  \\
&    SWIRE  &    S4\_4  &                   TYC3877-484-1  &      16:03:12.065  &     54:44:27.658  &       9.047  &       8.361  &       7.846  &       7.288  &      ...  &                       K2IIIFe-1  &       0.147  &                             ...  \\
&    SWIRE  &    S4\_5  &                        HD234308  &      16:26:05.554  &      52:18:08.266  &       8.652  &       8.101  &       7.932  &       7.407  &       K0  &                     K1IIIFe-0.5  &       0.237  &                             ...  \\
&      DGL  &    D4\_1  &                  TYC4419-1623-1  &     16:14:22.875  &     69:55:54.455  &      10.093  &       9.624  &       9.419  &       8.810  &      ...  &                             M2V  &       0.373  &                             ...  \\
&      DGL  &    D4\_2  &                  TYC4419-1631-1  &     16:18:10.929  &     69:16:36.761  &       9.923  &       9.466  &       9.229  &       8.916  &      ...  &                             K1V  &       0.124  &                             ...  \\
&      DGL  &    D4\_3  &                       BD+67-943  &     16:29:52.210  &     66:47:45.154  &       9.390  &       9.120  &       8.606  &       8.417  &       F8  &                            F8Ia  &       0.110  &                             ...  \\
&      DGL  &    D4\_4  &                  TYC4196-2280-1  &     16:34:34.354  &      65:36:05.818  &      10.424  &       9.946  &       9.783  &       9.339  &      ...  &                             G4V  &       0.232  &                             ...  \\
&      DGL  &    D4\_5  &                        HD151286  &     16:40:37.776  &     70:34:14.772  &       7.110  &       6.668  &       6.237  &       5.794  &      ...  &                            G3II  &       0.070  &                             ...  \\
&      DGL  &    D4\_6  &                       BD+69-873  &     16:47:31.365  &      68:51:02.603  &       8.338  &       7.820  &       7.495  &       7.010  &       K0  &                        G7.5IIIa  &       0.111  &                             ...  \\
&      DGL  &    D4\_7  &                        HD154273  &     16:58:40.137  &      69:38:05.431  &       7.022  &       6.508  &       6.197  &       5.746  &       K0  &                        G7.5IIIa  &       0.106  &                             ...  \\
&      DGL  &    D4\_8  &                  TYC4424-1380-1  &      17:08:33.058  &      71:00:28.044  &       9.242  &       8.911  &       9.008  &       8.727  &      ...  &                            G2IV  &       0.109  &                             ...  \\
&      DGL  &    D4\_9  &                  TYC4421-2278-1  &     17:16:54.688  &     67:38:26.279  &       8.993  &       8.460  &       8.269  &       7.792  &      ...  &                     K1IIIFe-0.5  &       0.174  &                             ...  \\
&  BootesB  &   BB4\_1  &                   TYC2557-870-1  &      14:40:08.540  &     34:40:29.669  &      10.107  &       9.545  &       9.249  &       8.768  &      ...  &                             M2V  &       0.331  &                             ...  \\
&  BootesB  &   BB4\_2  &                       HD128094  &     14:34:10.846  &     30:59:10.356  &       7.857  &       7.240  &       6.963  &       6.405  &       K0  &                           K2III  &       0.226  &                             ...  \\
&  BootesB  &   BB4\_3  &                   TYC2559-388-1  &     14:34:47.808  &      35:34:09.419  &       9.761  &       9.346  &       9.011  &       8.550  &      G8V  &                           G6III  &       0.184  &                             ...  \\
\multirow{3}{*}{4th} &  BootesB  &   BB4\_4  &                   TYC2553-947-1  &     14:28:52.868  &     31:30:30.316  &       8.505  &       7.763  &       7.642  &       6.917  &      ...  &                           K2III  &       0.170  &                             ...  \\
&  BootesB  &   BB4\_5  &                        V*KT-Boo  &      14:29:02.513  &     33:50:38.929  &       8.699  &       8.271  &       7.846  &       7.465  &        G  &                         G0Ib-II  &       0.074  &                             ...  \\
&  BootesB  &   BB4\_6\tablenotemark{h}  &                        HD126210  &     14:23:24.060  &     33:34:19.099  &       8.764  &       8.749  &       7.653  &       7.492  &       F8  &                            F1II  &       0.194  &                             ...  \\
&  BootesB  &   BB4\_7  &                   TYC2549-413-1  &     14:23:23.452  &     34:33:24.854  &       9.399  &       8.885  &       8.510  &       7.947  &      ...  &                  K1IIIbCN1.5Ca1  &       0.269  &                             ...  \\
	\enddata
		\tablenotetext{a,b}{The J2000.0 right ascension (RA) and the declination (Dec) of a star in a sexagesimal from 2MASS data.}
		\tablenotetext{c,d}{Vega magnitude of the LRS.}
		\tablenotetext{e,f}{Vega magnitude of the matched 2MASS point source catalog.}
		\tablenotetext{g}{Spectral type given from SIMBAD database.}
		\tablenotetext{h,i,j,k,l}{A star that observed from two independent flights.}
	\end{deluxetable}
\end{landscape}

\clearpage


\begin{thebibliography}{}
\bibitem[Arai et al.(2015)]{arai15} Arai, T., Matsuura, S., Bock, J., et al. \ 2015, \apj, 806, 69
\bibitem[Bock et al.(2006)]{bock06} Bock, J., Battle, J., Cooray, A., et al. \ 2006, New A Rev., 50, 215
\bibitem[Bock et al.(2013)]{bock13} Bock, J., Sullivan, I., Arai, T., et al. 2013, \apjs, 207, 32
\bibitem[Bohlin \& Gilliland (2004)]{bohlin04} Bohlin, R. C., \& Gilliland, R. L., 2004, \aj, 127, 3508 (BG)
\bibitem[Cohen et al. (2003)]{cohen03} Cohen, M., Wheaton, Wm. A., \& Megeath, S. T. 2003, \aj, 126, 1090
\bibitem[Cushing(2005)]{cushing05} Cushing, M. C. 2005, \apj, 623, 1115
\bibitem[Fischer et al. (1998)]{fischer98} Fisher, J., Baumback, M. M., Bowles, J. H., Grossmann, J. M., \& Antoniades, J. A. 1998, Proc SPIE, 3438, 23
\bibitem[Garnett \& Forrest (1993)]{gf93} Garnett, J., D.P., \& Forrest, W.J., 1993, SPIE, 1946, 395G
\bibitem[Gliese(1971)]{gliese71} Gliese, W. 1971, Veroeffentlichungen des Astronomischen Rechen-Instituts Heidelberg, 24, 1
\bibitem[Hauser \& Dwek (2001)]{hauser01} Hauser, M.G. \& Dwek, E., 2001, ARA\&A, 39, 249
\bibitem[Houk et al.(1999)] {houk99} Houk, N., \& Swift, C. 1999, University of Michigan Catalogue of Two-Dimensional Spectral Types for the HD Stars, Vol. 5 (Ann Arbor: Univ. Michigan)
\bibitem[Jaschek, M.(1978)]{jas78} Jaschek, M. \ 1978, CDS Inf. Bull. 15, 121
\bibitem[Jaschek \& Jaschek (1973)]{jaschek73} Jaschek, C., \& Jaschek, M. 1973, in IAU Symp. 50, Spectral Classification and Multicolour Photometry, ed. C. Fehrenbach \& B. E. Westerlund (Dordrecht:Reidel), 43
\bibitem[Jordi et al. (2010)]{jordi10} Jordi, C., Gebran, M., Carrasco, J. M., et al. 2010, A\&A, 523, A48
\bibitem[Joyce et al.(1998)]{joyce98} Joyce, R. R., Hinkle, K. H., Wallace, L., Dulick, M., \& Lambert, D. L., 1998, \aj, 116, 2520
\bibitem[Kessler et al. (1996)]{kessler96} Kessler, M. F., Steinz, J. A., \& Anderegg, M. E., et al. 1996, A\&A, 315, L27
\bibitem[Korngut et al.(2013)]{korngut13} Korngut, P. M., Renbarger, T., Arai, T., et al. 2013, \apjs, 207, 34
\bibitem[Lee et al.(2010)]{lee10} Lee, D. H., Kim, M. G., Tsumura, K., et al. \ 2010, Journal of Astronomy and Space Sciences, 27, 401
\bibitem[Leinert et al. (1998)]{leinert98} Leinert, Ch., Bowyer, S., Haikala, L. K., et al. 1998,  A\&AS, 127, 1L
\bibitem[Madau \& Pozzetti (2000)]{madau00} Madau, P. \& Pozzetti, L., 2000,  MNRAS, 312, L9-L15
\bibitem[Matsumoto et al.(2005)]{matsumoto05}Matsumoto, T., Matsuura, S., Murakami, H., et al. 2005, \apj, 626, 31
\bibitem[Matsuura et al.(1999)]{matsuura99} Matsuura, M., Yamamura, I., Murakami, H., Freund, M. M., \& Tanaka, M. 1999, A\&A, 348, 579
\bibitem[Matsuura et al. 2016, ApJ, submitted (2016)]{matsuura16} Matsuura, S., Arai, T., Bock, J., et al. 2016, ApJ, submitted
\bibitem[Meyer et al.(1998)]{meyer98} Meyer, M. R., Edwards, S., Hinkle, K. H., \& Strom, S. E., 1998, \apj, 508, 397
\bibitem[Murakami et al.(1996)]{murakami96} Murakami, H., Freund, M. M., Ganga, K., et al. 1996, \pasj, 48, L41
\bibitem[Peterson et al.(2008)]{peterson08} Peterson, D. E., Megeath, S. T., Luhman, K. L., et al. \ 2008, \apj, 685, 313
\bibitem[Perryman et al. (2001)]{perryman01} Perryman, M. A. C., de Boer, K. S., Gilmore, G., et al. 2001, A\&A, 369, 339
\bibitem[Pickles(1998)]{pickles98} Pickles, A. J. 1998, \pasp, 110, 863
\bibitem[Rayner(2009)]{rayner09} Rayner, J. T. 2009, \apjs, 185, 289
\bibitem[Roeser(1988)]{roeser88} Roeser, S. \& Bastian, U., 1988, A\&A, 74, 449
\bibitem[Russell(1934)]{russell34} Russell, H. N. 1934, \apj, 79, 317
\bibitem[Schlegel(1998)]{schlegel98} Schlegel, D. J. 1998, \apj, 500, 525
\bibitem[Skrutskie et al.(2006)]{skrutskie06} Skrutskie, M. F., Cutri, R. M., Stiening, R., et al. \ 2006, \aj, 131, 1163S
\bibitem[Sorahana \& Yamamura (2014)]{sora14} Sorahana, S., Yamamura, I. \ 2014, \apj, 793, 47
\bibitem[Tsumura et al.(2010)]{tsumura10} Tsumura, K., Battle, J., Bock, J., et al. \ 2010, \apj, 719, 394
\bibitem[Tsumura et al.(2013)]{tsumura13} Tsumura, K., Arai, T., Battle, J., et al. \ 2013, \apjs, 207, 33
\bibitem[Tsuji et al. (1997)]{tsuji97} Tsuji, T., Ohnaka, K., Aoki, W., \& Yamamura, I. 1997, A\&A, 320, L1
\bibitem[Tsuji et al. (2000)]{tsuji00} Tsuji, T. 2000, \apj, 538, 801
\bibitem[Tsuji et al. (2001)]{tsuji01} Tsuji, T. 2001, A\&A, 376, L1
\bibitem[Tsuji et al. (2015)]{tsuji15} Tsuji, T. 2015, PASJ, 67, 26T
\bibitem[Wenger et al. (2000)]{wenger00} Wenger, M., Ochsenbein, F., Egret, D., et al. 2000, A\&AS, 143, 9
\bibitem[Wing \& Spinrad (1970)]{ws70} Wing, R. F., \& Spinrad, H. 1970, \apj, 159, 973
\bibitem[Woolf et al. (1964)]{woolf64} Woolf, N. J., Schwarzschild, M., \& Rose, W. K. 1964, \apj, 140, 833
\bibitem[Yuan et al. (2013)]{yuan13} Yuan, H. B., Liu, X. W., \& Xiang, M. S. 2013, \mnras, 430, 2188
\bibitem[Zemcov et al.(2013)]{zemcov13} Zemcov, M., Arai, T., Battle, J., et al. 2013, \apjs, 207, 31


\end{thebibliography}
\end{document}